\documentclass[prd,twocolumn]{revtex4}
\usepackage{graphicx,natbib}
\usepackage{url}
\usepackage{color}
\usepackage{rotating}
\usepackage{amssymb,amsmath}

\newcommand{\apjl}{{Astrophys.~J.~Lett.}}
\newcommand{\apjs}{{Astrophys.~J.~Supp.}}

\newcommand{\Mobs}{M^{\rm obs}}
\newcommand{\Mbias}{M^{\rm bias}}
\newcommand{\Mth}{M^{\rm th}}
\newcommand{\siglnM}{\sigma_{\ln M}}
\newcommand{\zphot}{z^{\rm p}}
\newcommand{\zbias}{z^{\rm bias}}
\newcommand{\sigz}{\sigma_{z}}
\newcommand{\Msun}{M_{\odot}}
\newcommand{\DE}{\Omega_{\rm DE}}
\newcommand{\varsz}{\sigma^{2}_{\rm sz}}
\newcommand{\varopt}{\sigma^{2}_{\rm opt}}
\begin{document}

\title{Cross-calibration of cluster mass-observables}
\author{Carlos Cunha}
\email{ccunha@umich.edu}
\affiliation{%
Department of Astronomy and Astrophysics, University of Chicago, Chicago, IL 60637 \\
Kavli Institute for Cosmological Physics, University of Chicago, Chicago, IL 60637 \\
Department of Physics, University of Michigan, Ann Arbor, MI 48109
}%

\date{\today}

 \begin{abstract}
This paper is a first step towards developing a formalism to optimally
extract dark energy information from number counts using multiple
cluster observation techniques.
We use a Fisher matrix analysis to study the improvements in 
the joint dark energy and cluster mass-observables constraints
resulting from combining cluster counts and clustering abundances 
measured with different techniques.
We use our formalism to forecast the constraints in $\Omega_{\rm DE}$ 
and $w$ from combining optical and SZ cluster counting
on a 4000 sq. degree patch of sky.
We find that this cross-calibration approach yields $\sim 2$ times better 
constraints on $\Omega_{\rm DE}$ and $w$ compared to simply adding the Fisher 
matrices of the individually self-calibrated counts.
The cross-calibrated constraints are less sensitive to variations in the mass threshold
or maximum redshift range.
A by-product of our technique is that the correlation
between different mass-observables is well constrained without the need 
of additional  priors on its value.
 \end{abstract}

\maketitle
\section{Introduction}

The evolution of the number of clusters of galaxies provides a powerful tool to
study the nature of dark energy.
Clusters are sensitive probes of the growth of structure because cluster
abundances are exponentially dependent on the linear density perturbation 
field.
In addition, cluster surveys are sensitive to the evolution of the volume
element with redshift so that cluster surveys also probe the background
cosmology.

Planned and ongoing cluster surveys will detect millions of 
clusters using a variety of techniques such as counts of optically 
detected galaxies (e.g. DES \cite{des05}, LSST, \cite{tys02}), the Sunyaev-Zel'dovich (SZ)
flux decrement (e.g. SPT \cite{ruh04} and ACT \cite{kos03}), 
X-ray temperature and surface brightness (e.g. eRosita, \cite{pre07}),
and weak lensing shear.
Because different cluster techniques suffer from different sources of 
errors, combining the information from different
surveys is essential to reduce random errors and control the systematics.

One of the major challenges in extracting dark energy information from
clusters is that cluster masses are not directly observable.
One must rely on observable proxies for mass which only correlate 
statistically with the true mass. 
The inherent uncertainties in the observable-mass relation will 
degrade cosmological constraints if not well understood.
Methods have been developed to use additional cluster properties
such as the cluster power spectrum \citep{maj04}, sample covariance
from counts in cells \citep{lim04}, or the shape of the observed
mass function \citep{lim05,hu03,roz07,lev02} to ``self-calibrate'' the 
mass-observable relation by simultaneously solving for the cosmological
and mass-observable parameters.

Other works have investigated combining different cluster techniques 
to cross-calibrate the mass-observable relations of each \citep{maj04,maj03,you06}.
In \cite{maj04,maj03}, the cross-calibration is between an SZ or X-ray survey
and a detailed mass follow-up to calibrate the mass-observable relation, whereas 
\cite{you06} combine SZ and X-ray surveys.
However, these studies have assumed that the two surveys 
were independent, so that the joint constraints were estimated by 
adding the Fisher matrices of both experiments.
But if two surveys observe the same patch of sky, the measurements
are {\bf not} independent.
The goal of this paper is to show how to exploit the interdependence
of cluster surveys over the same patch of sky to improve constraints
on dark energy and mass-nuisance parameters.

The paper is organized as follows.
In \S \ref{sec:selfcal} we describe the Fisher matrix formalism to 
forecast cosmological constraints from cluster counts and clustering using 
a single and multiple observables.
We describe the major cluster mass determination techniques in 
\S \ref{sec:techniques} and explain our parametrization of the errors in 
the observables, i.e. the mass-observable distributions.
Results are presented in \S \ref{sec:res} and 
our conclusions and prospects for future work are given in \S \ref{sec:conc}.

\section{Self-calibration and the Fisher matrix formalism}\label{sec:selfcal}

In this section we review how to obtain cosmological constraints from 
cluster counts and clustering using a single or multiple observables.
Combining counts and clustering to derive cosmological constraints from 
a single mass estimation technique is often referred to as self-calibration.

\subsection{Mean number counts}\label{sec:counts}

The use of clusters of galaxies as cosmological indicators depends on how reliably
N-body simulations can predict the number density of dark matter halos associated
to clusters of a given mass given an initial power spectrum.
We adopt the fitting function of \cite{jen01} for differential
comoving number density of clusters
\begin{equation}
\frac{d \bar n}{d\ln M} = 0.3 \frac{\rho_{m}}{M} \frac{d \ln \sigma^{-1}}{d\ln M}
        \exp[-|\ln \sigma^{-1} + 0.64|^{3.82}],
\label{eqn:mfunc}
\end{equation}
\noindent where $\sigma^2(M,z)$ is the variance of the density field in a 
spherical region with mean (present-day) matter density $\rho_m$ encircling a mass $M$.
Even though more recent fitting-functions exist (e.g. \cite{tin08,war06}),
we adopt the above for easier comparison with the literature 
(e.g. \cite{lim07,lim05,lim04}) and because the results are relatively insensitive 
to the fiducial mass function used.

Eq. (\ref{eqn:mfunc}) shows that the number density of clusters is 
sensitive to the variance of the density field, and hence to the initial power spectrum.
However, uncertainties in the estimation of the mass
 are degenerate with changes in cosmological parameters.
The utility of cluster number counts is therefore limited by uncertainties in the
mass-observable relation.
Results from both simulations (e.g. \cite{sha07,kra06}) and observations 
(e.g. \citep{ryk08b,evr08,ryk08a}) suggest that the mass-observable relations
can be parametrized in simple forms with lognormal scatter of the mass-observable
about the mean relation.
Other works (see e.g. \cite{coh07}) suggest that the distribution of galaxies
in halos may be more complicated. 

For $n$ observables, the probability of measuring clusters given the true mass $M$ 
and redshift $z$ is 
\begin{equation}
p({\bf \Mobs},\zphot|M,z)\phi({\bf \Mobs},\zphot),
\end{equation}
\noindent where ${\bf \Mobs}=({\Mobs_1,\Mobs_2,...,\Mobs_n})$ and $\phi(\Mobs)$ 
is the combined selection function for all the observables.
For simplicity, we always work in a range of redshift and mass where the 
surveys are expected to be nearly complete.
This allows us to approximate the selection function as unity.
This range depends on the observable we are using, so we postpone 
justifying our assumptions for survey selections to \S \ref{sec:techniques},
when we describe the different cluster techniques. 
We further assume that the redshift errors are independent of the mass-observable
errors.
This assumption is not strictly true, since the bigger the cluster, the more 
bright optical galaxies it should have, and the better the cluster 
redshift estimate will be.
This is particularly relevant for optical clusters, for which the cluster detection 
and mass estimate are inseparable from the cluster redshift determination.
We will postpone dealing with this difficulty to a later work. 
For now, we write
\begin{equation}
p({\bf \Mobs},\zphot|M,z)=p({\bf \Mobs}| M)p(\zphot|z)
\end{equation}

We define the probability of measuring the observable $\Mobs$ given
the true mass $M$ as \citep{lim05}

\begin{equation}
p(\Mobs | M) = \frac{1}{ \sqrt{2\pi \siglnM^2}}  \exp\left[ -x^2(\Mobs) \right] ,
\end{equation}

\noindent where

\begin{equation}
x(\Mobs) \equiv \frac{ \ln \Mobs - \ln M - \ln  \Mbias(M,z)}{ \sqrt{2 \siglnM(M,z)^2}} .
\label{eqn:x1mobs}
\end{equation}
\noindent We describe our parametrization of $\Mbias(M,z)$ and $\siglnM(M,z)^2$ 
in \S \ref{sec:techniques} when we discuss our modeling of different cluster techniques.

The number density of clusters at a given redshift $z$ with observable in
the range $\Mobs_{\alpha} \leq \Mobs \leq \Mobs_{\alpha+1}$ is given by

\begin{eqnarray}
\bar n_{\alpha}(z) &\equiv& \int_{\Mobs_{\alpha}}^{\Mobs_{\alpha+1}} \frac{d \Mobs}{\Mobs}
\int {\frac{dM}{M}} { \frac{d \bar n}{d\ln M}}
p(\Mobs | M)
\label{eqn:nofz}
\end{eqnarray}
\noindent where $x_{\alpha}=x(\Mobs_{\alpha})$.

We define the probability of measuring two observables $\Mobs_a$, $\Mobs_b$ 
given the true mass as a bivariate Gaussian distribution
\begin{equation}
p(\Mobs_1,\Mobs_2|M)=\frac{1}{(2\pi)\det({\bf C})^{1/2}} \exp\left[ -\frac{{\bf x}^{\rm T} {\bf C}^{-1} {\bf x}}{2}\right]
\end{equation}
\noindent where $\bf C$ is the covariance matrix defined as
\begin{equation}
{\bf C}=\begin{pmatrix}
{\sigma}_{a}^2 & \rho {\sigma}_{a}{\sigma}_{b}\\
\rho {\sigma}_{a}{\sigma}_{b} & {\sigma}_{b}^2
\end{pmatrix} \label{eqn:covscat}
\end{equation}
\noindent and $\rho \in [-1,1]$ is the correlation coefficient. 
We motivate the use of the bivariate distribution in Appendix \ref{appendix}.

At a given redshift $z$, the average number density of clusters with observables such that
$\Mobs_{a,\alpha} \leq \Mobs_{a} \leq \Mobs_{a,\alpha+1}$ and $\Mobs_{b,\beta} \leq \Mobs_{b} \leq \Mobs_{b,\beta+1}$ is given by

\begin{widetext}
\begin{eqnarray}
 \bar n_{\alpha, \beta}(z)  &\equiv& \int_{\Mobs_{a,\alpha}}^{\Mobs_{a, \alpha+1}} \frac{d \Mobs_a}{ \Mobs_a} 
\int_{\Mobs_{b,\beta}}^{\Mobs_{b, \beta+1}} \frac{d \Mobs_b}{\Mobs_b}
 \int \frac{dM}{M} \frac{ d \bar n}{d\ln M} p(\Mobs_a, \Mobs_b | M) \nonumber \\
&=&\frac{\sqrt{\pi}}{8} \int \frac{d M}{ M} \frac{ d \bar n }{ d\ln M} \int_{\Mobs_{a,\alpha}}^{\Mobs_{a, \alpha+1}} \frac{d \Mobs_a}{ \Mobs_a} e^{-{x}_{a}^2} \left[{\rm erfc}  \left(\rho x_{a}- x_{b}(\Mobs_{b,\beta}) \over \sqrt{ (1-\rho^2)} \right) - {\rm erfc}\left(\rho x_{a}- x_{b}(\Mobs_{b,\beta+1}) \over \sqrt{ (1-\rho^2)} \right)   \right] \label{eqn:nofz2mobs} 
\end{eqnarray}
\end{widetext}
\noindent
For the two observables case, the integrals over the observables can only be 
performed analytically if $\rho=0$.
One would think that this problem could be resolved by diagonalizing 
the inverse covariance matrix - defined in Eq. (\ref{eqn:covscat}). 
Diagonalization, however, does not simplify the calculation because the limits of the innermost 
integral over observables become dependent on the other observable. 
Thus, one cannot avoid performing the numerical integration.
The equation for $b(z)$ is modified analogously to Eq. (\ref{eqn:nofz2mobs}).

We interpret Eq. (\ref{eqn:nofz2mobs}) as the combination of 
the error-free number density multiplied by two window-functions defined as:
\begin{equation}
W^e_1=e^{-{x}_{a}^2 \over 2}
\label{eqn:ewin1}
\end{equation}
\noindent and
\begin{eqnarray}
W^e_2&=&{\rm erfc} \left(\frac{\rho x_{a}- x_{b}(\Mobs_{b,\beta})}{\sqrt{2 (1-\rho^2)}} \right) \nonumber \\
&& - {\rm erfc}\left(\frac{\rho x_{a}- x_{b}(\Mobs_{b,\beta+1})}{\sqrt{2 (1-\rho^2)}} \right).
\label{eqn:ewin2}
\end{eqnarray}
Window $W^e_1$ has characteristic width given by the scatter of the observable 
$a$ with respect to the true mass, and is centered, in the $\ln \Mobs_{a} - \ln M$ 
coordinate, at the bias in the mass-observable relation, $\ln\Mbias_{a}$. 
The shape and position of window $W^e_2$ in  $(\ln \Mobs_{a} - \ln M)$ 
depend on the value of the correlation coefficient $\rho$ as well as on the boundaries of
the mass bin of the observable $b$, $\Mobs_{b,\beta}$ and $\Mobs_{b, \beta+1}$.
If $\rho=0$, $W^e_2$ is simply a constant, independent of $\Mobs_{a}$ and $M$, as expected. 
For finite $\rho$, $W^e_2$ has the shape of a Mexican hat. 
As $|\rho| \rightarrow 1$, $W^e_2$ approaches a top-hat function, with edges at 
$x_{b}(\Mobs_{b,\beta})$ and $x_{b}(\Mobs_{b,\beta+1})$ for positive $\rho$ or at 
$-x_{b}(\Mobs_{b,\beta+1})$ and $-x_{b}(\Mobs_{b,\beta})$ for negative $\rho$.
$W^e_2$ is {\bf not} invariant under $\rho \rightarrow -\rho$ transformations.
Decreasing $\rho$ ``spreads out'' the number counts in the $\Mobs_a - \Mobs_b$ plane.
If the observables have different scatter, the spreading will be 
asymmetric with respect to the $\Mobs_a=\Mobs_b$ line.
In other words, variations in $\rho$ are partially degenerate with both the scatter and bias of the 
different observables.

The mean cluster number counts are given by integrating Eq. (\ref{eqn:nofz}) 
or (Eq. \ref{eqn:nofz2mobs}) over comoving volume. 
In spherical comoving coordinates, the volume element $dV$ is 

\begin{eqnarray}
dV =  r^2 dr d\Omega =\frac{r^2(z)}{H(z)}dz d\Omega ,
\label{eqn:vofz}
\end{eqnarray}
\noindent where $H(z)$ is the Hubble parameter at redshift $z$, $r(z)$ 
is the comoving angular diameter distance and $d\Omega$ is the differential
solid angle.
Uncertainties in the redshifts distort the volume element.
Assuming photometric techniques are used to determine the redshifts of 
the clusters, we parametrize the probability of measuring a photometric
redshift, $\zphot$, given the true cluster redshift $z$ as \cite{lim07}

\begin{eqnarray}
p(\zphot| z) &=& {1 \over \sqrt{2\pi \sigz^2  } } \exp\left[ -y^2(\zphot) \right] ,
\label{eqn:pzpzs}
\end{eqnarray}

\noindent where

\begin{eqnarray}
y(\zphot)&\equiv& \frac{\zphot -z -\zbias}{ \sqrt{2\sigz^2}} 
\end{eqnarray}

\noindent and $\zbias=\zbias(z)$ is the photometric redshift bias and $\sigz^2=\sigz^2(z)$
is the variance in the photo-z's.
We parametrize them as
\begin{eqnarray}
\zbias(z)\equiv \zbias_0 + d_1(1+z) \\
\sigz(z)\equiv \sigma^{0}_{z} + e_1(1+z) 
\end{eqnarray}
\noindent For this paper we set the fiducial values $\zbias_0=d_1=e_1=0$, and $\sigma^{0}_{z}=0.02$,
the expected overall scatter of cluster photo-z's in the Dark Energy Survey \citep{des05}.
We hold these parameters fixed throughout.

Assuming perfect angular selection the mean number of clusters 
in a photo-z bin $\zphot_i \le \zphot \le \zphot_{i+1}$ is

\begin{eqnarray}
\bar m_{\alpha,\beta,i} &=& \int_{\zphot_i}^{\zphot_{i+1}} d\zphot 
\int dV \bar n_{\alpha,\beta} W_{i}^{\rm th}(\Omega) p(\zphot| z)
\label{eqn:numwin}
\end{eqnarray}

\noindent where $W^{\rm th}_{i}(\Omega)$ is an angular top hat window function.  

To simplify the notation, henceforth we use the index $\alpha$ to 
indicate bins of both observables.

\subsection{Noise in counts}\label{sec:sij}

The number of clusters found in an angular/redshift bin can deviate 
from the mean counts because of Poisson noise and large scale 
structure clustering.
Both effects must be included in any likelihood analysis.
On cluster scales, the clustering of baryonic matter follows the linear density
fluctuations of total matter $\delta(x)$ corrected by the linear bias.
That is,

\begin{eqnarray}
m_{\alpha,i}({\bf x}) &=& \bar m_{\alpha,i}[ 1 + b_{\alpha,i}(z) \delta({\bf x}) ] ,
\label{eqn:countsbias}
\end{eqnarray}

\noindent where $b_{\alpha,i}(z)$ is the 
average cluster linear bias defined as

\begin{eqnarray}
b_{\alpha,i}(z) &=& \frac{1}{\bar n_{\alpha,i}(z)}  \int \frac{d{\Mobs_\alpha}}{\Mobs_\alpha}\int \frac{d{\Mobs_\beta}}{\Mobs_\beta} \int \frac{d M}{M} \nonumber \\
&&\times \frac{d \bar n_{\alpha,i}(z)}{d\ln M} b(M;z)p(\Mobs|M).
\end{eqnarray}

\noindent We adopt the $b(M;z)$ fit of \cite{she99}:

\begin{equation}
b(M;z) = 1 + \frac{a_c \delta_c^2/\sigma^2 -1}{\delta_c} 
         + \frac{ 2 p_c}{\delta_c [ 1 + (a \delta_c^2/\sigma^2)^{p_c}]}
\label{eqn:biasofmz}
\end{equation}

\noindent with $a_c=0.75$,  $p_c= 0.3$,  and $\delta_c=1.69$.

The sample covariance of counts $m_{\alpha,i}$ is, given by \citep{hu03}

\begin{eqnarray}
S^{\alpha \beta}_{ij} &=&\langle (m_{\alpha,i} -\bar m_{\alpha,i})(m_{\beta,j} - \bar m_{\beta,j})\rangle \label{eqn:sija} \\
&=& b_{\alpha,i} \bar m_{\alpha,i} b_{\beta,j} \bar m_{\beta,j} \nonumber \\
&&\times \int{d^3 k \over (2\pi)^3} W_i^*({\bf k})W_j({\bf k})\sqrt{P_i(k)P_j(k)}, \label{eqn:sijb}
\end{eqnarray}
\noindent where $W_i^*(\bf k)$ is the Fourier transform of the top-hat 
window function and $P_i(k)$ is the linear power spectrum at the centroid 
of redshift bin $i$.
Notice that, in contrast to \cite{hu03}, we use $\sqrt{P_i(k)P_j(k)}$ 
instead of $P(k)$ at an average redshift.
We do not notice significant differences from this change. 
In addition, for computational efficiency, we only calculate covariance terms for
which $|i-j| \leq 1$
and set the remaining terms to zero.
Going from Eq. (\ref{eqn:sija}) to Eq. (\ref{eqn:sijb}) we assumed that 
the bias was approximately constant in each photo-z bin so that it could 
be removed from the integral.
We only considered the sample covariance in bins of redshift, but the angular covariance
also contains useful information.
We postpone calculating the full sample covariance to a future work.

Following \cite{lim07}, we find that the window function  $W_i^*(\bf k)$ 
in the presence of photo-z errors is given by

\begin{eqnarray}
W_i(\bf k)=&&2\exp{\left[ i k_{\parallel} \left( r_i+\frac{ {\zbias_i} }{ {H_i} } \right) \right] }
              \exp{ \left[- \frac{{\sigma_{z,i}^2 k_{\parallel}^2}}{{2H_i^2}}  \right] } \nonumber \\
           && \times  \frac{\sin( k_{\parallel} \delta r_i/2) }{k_{\parallel} \delta r_i/2} 
              \frac{J_1(k_{\perp} r_i \theta_s)}{k_{\perp} r_i \theta_s}. 
\label{eqn:window}
\end{eqnarray}
\noindent Here $r_i=r(\zphot_i)$ is the angular diameter distance to the 
$i^{\rm th}$ photo-z bin, and $\delta r_i=r(\zphot_{i+1})-r(\zphot_{i})$.
Similarly, $H_i=H(\zphot_i)=H(z)$, $\zbias_i=\zbias(\zphot_i)=\zbias(z)$, and 
$\sigma_{z,i}=\sigma_z(\zphot_i)=\sigma_z(z)$.
We assumed that $H(z)$, $\zbias(z)$, and $\sigma_z(z)$ are constant inside each bin.

The Poisson noise of the counts is fully specified by the mean counts $\bar m$.
The sample variance in the counts is determined by the mean counts, 
the bias, and the initial power spectrum.
Since all these quantities can be predicted theoretically, both the mean counts
 and the sample variance contain useful information.
In the following section we use the Fisher matrix formalism to estimate joint 
constraints for dark energy and mass-observable parameters using the 
information in the counts and the noise.

\subsection{Fisher Matrix}\label{sec:fishmat}

Given a model specified by a set of parameters $p_{\alpha}$, with likelihood $L$, the Fisher information
matrix is defined as
\begin{equation}
F_{\alpha \beta}=-\left \langle \frac{\partial^2 {\rm ln} L}{\partial p_\alpha \partial p_\beta}\right \rangle
\end{equation}

The marginalized errors in the parameters are given 
by $\sigma(p_\alpha)=\left[(F^{-1})_{\alpha \alpha} \right]^{1/2}$.
Priors are easily incorporated into the Fisher matrix.
If parameter $p_{i}$ has a prior uncertainty of $\sigma(p_{i})$, we simply 
add $\sigma(p_{i})^{-2}$ to the $F_{ii}$
entry of the Fisher matrix before inverting.

Define the covariance matrix
\begin{equation}
C_{ij} = S_{ij} + {\bar m_i} \delta_{ij}
\label{eqn:covdat}
\end{equation}
\noindent where ${\bar m_i}$ is the vector of mean counts defined in Eq. (\ref{eqn:numwin}) and 
$S_{ij}$ is the sample covariance defined in Eq. (\ref{eqn:sijb}).
The indices $i$ and $j$ here run over all mass and redshift bins.
Assuming Poisson noise and sample variance are the only sources of 
noise, the Fisher matrix is, \citep{hu06,lim04,hol01}

\begin{equation}
F_{\alpha\beta}=  \bar{\bf m}^t_{,\alpha} {\bf C}^{-1}
 \bar{\bf m}_{,\beta} 
+ {1\over 2} {\rm Tr} [{\bf C}^{-1} {\bf S}_{,\alpha}
 {\bf C}^{-1} {\bf S}_{,\beta} ],
 \label{eqn:fishmat}
\end{equation}
where the ``,'' denote derivatives with respect to the model parameters.
The first term on the right-hand side contains the ``information'' 
from the mean counts, $\bar m$.
The $S_{ij}$ matrix only contributes noise to this term, and hence only reduces 
its information content.
The second term contains the information from the sample covariance.

For our purposes, the model parameters are the cosmological 
parameters, the parameters describing the errors in the observables 
(i.e. the mass nuisance parameters), and the parameters of the photo-z errors.
We use two sets of fiducial cosmological parameters.
One set is based on the first year data release of the Wilkinson Microwave 
Anisotropy Probe (WMAP1, \cite{ben03}) and the other is based on 
the third-year data release (WMAP3, \citep{spe07}).
We use WMAP1 and WMAP3 instead of the more recent five-year data 
release because the WMAP1 and WMAP3 are more extreme cases with regards to 
the value of $\sigma_8$ and the predicted number counts, and hence WMAP5 is 
more or less in-between both of them.
The WMAP1 parameters assumed are: the baryon density, $\Omega_b h^2=0.024$,
the dark matter density, $\Omega_m h^2 =0.14$, the normalization of the power 
spectrum at $k=0.05 {\rm Mpc}^{-1}$, $\delta_{\zeta}=5.07 \times 10^{-5}$, the
tilt, $n=1.0$, the optical depth to reionization, $\tau=0.17$, the dark energy density, 
$\Omega_{\rm DE}=0.73$, and the dark energy equation of state, $w=-1$.
In this cosmology, $\sigma_8=0.91$.
For WMAP3 we set  $\Omega_b h^2=0.0223$, $\Omega_m h^2 =0.128$,  
$\delta_{\zeta}=4.053 \times 10^{-5}$ at $k=0.05 {\rm Mpc}^{-1}$, $n=0.958$, 
$\tau=0.093$, $\Omega_{\rm DE}=0.73$, and $w=-1$.
This cosmology corresponds to $\sigma_8=0.76$.
With the exception of $w$, the cosmological parameters we used have been determined to 
an accuracy of a few percent.
Extrapolating into the future, we assume $1\%$ priors on all cosmological parameters 
except $\Omega_{\rm DE}$ and $w$.
We used CMBfast \citep{sel96}, version 4.5.1, to calculate the transfer functions.

\section{Cluster mass determination techniques}\label{sec:techniques}

 \begin{figure*}
  \begin{minipage}[t]{70mm}
    \begin{center}
      \resizebox{70mm}{!}{\includegraphics[angle=0]{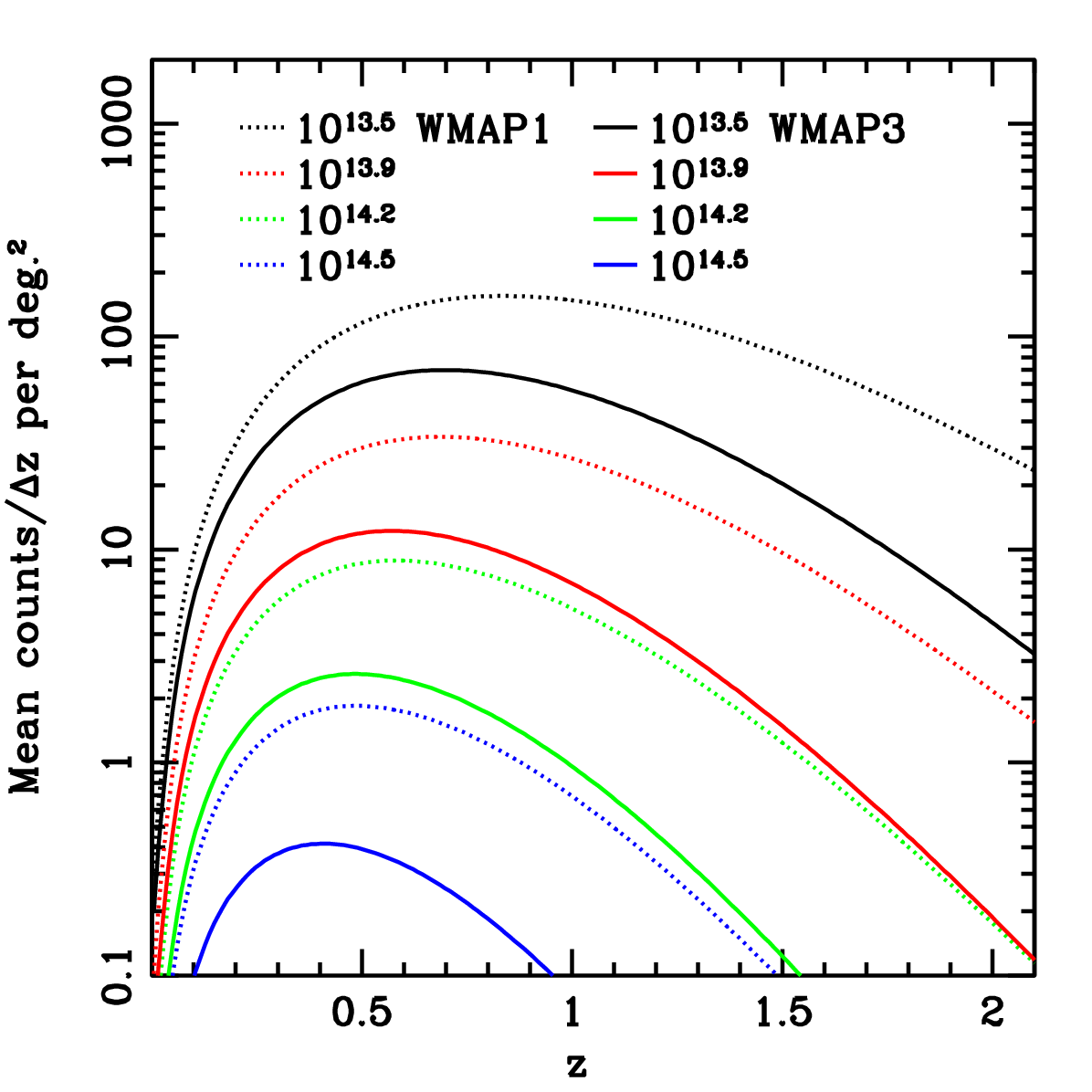}}
    \end{center}
  \end{minipage}
  \begin{minipage}[t]{70mm}
    \begin{center}
      \resizebox{70mm}{!}{\includegraphics[angle=0]{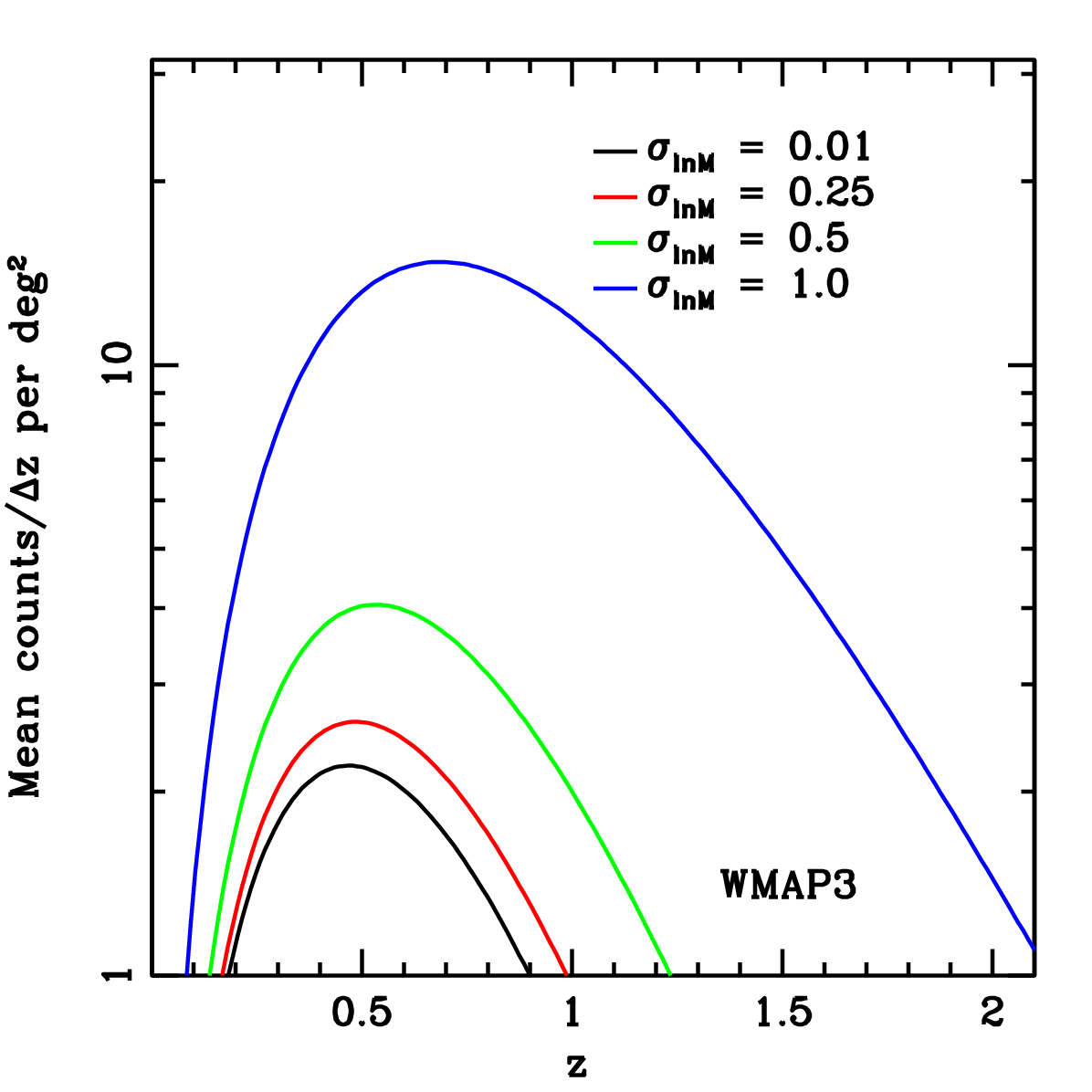}}
    \end{center}
  \end{minipage}
  \caption{({\it Left}) Mean counts as a function of redshift $\bar m(z)$ for various mass thresholds, with $\siglnM=0.25$ for 
both WMAP1 and WMAP3 cosmologies.
    ({\it Right}) $\bar m(z)$ for various values of $\siglnM$, with $\Mth=10^{14.2}h^{-1}\Msun$ 
assuming a WMAP3 cosmology.
  }\label{fig:counts.z}
  \end{figure*}

There are four commonly used cluster detection techniques for which large surveys are planned:
optical, X-ray, Sunyaev-Zeldovich flux decrement, and weak lensing.
For our Fisher matrix purposes, each of them is fully specified by a mass threshold, 
survey area, maximum redshift, and the parameters for the fiducial errors in $\Mobs$ 
and $\zphot$.

We show the mean number counts per redshift
bin per sq. degree as a function of photometric redshift (with a constant scatter 
of $\sigma_{z}^0=0.02$) for several mass thresholds and scatters in Fig \ref{fig:counts.z}.
The {\it left} plot shows the mean counts for $\Mth=10^{13.5}$, $10^{13.9}$, $10^{14.2}$, 
and  $10^{14.2} h^{-1}\Msun$, for a fixed scatter of $\siglnM=0.25$.
The sensitivity of the counts to the mass threshold is apparent. 
The plot on the right shows the mean counts for $\siglnM=0.01$, $0.25$, $0.5$, $1.0$ with the
threshold set to $\Mth=10^{14.2}h^{-1} \Msun$. 
The increase of the scatter results in an increase in the total counts because the mass
function falls exponentially with mass. 
It also causes flattening of the $\bar m(z)$ curve.
The increase in the scatter implies an increase in the variance in counts, but a 
decrease in the shot noise. 
For {\it perfectly known scatter}, the decrease in shot noise outweighs the 
increase in variance implying that more scatter can yield better cosmological constraints.
However, it is harder to constrain larger scatter and its evolution, and the assumption of 
Gaussianity may break down.
This issue is particularly relevant for a WMAP3 cosmology, where there are 
fewer clusters compared to WMAP1.

Since the focus of this paper is on combining clusters in the same area of the sky,
we limit our tests to surveys overlapping the South Pole Telescope (SPT) SZ Cluster survey.
We thus set the area of the sky to 4000 square degrees, which we subdivide into 400 bins 
of 10 sq. degrees each.
We assume SPT will be able to observe clusters with $\Mobs \geq 10^{14.2} h^{-1} \Msun$ 
up to a redshift of 2 (see e.g. \cite{car02}).
We assume that photometric redshifts will be available using DES+VISTA photometry.
We parametrize the SZ mass bias and variance as
\begin{eqnarray}
{\rm ln}\Mbias(M,z)&=&{\rm ln}\Mbias_0 + a_1(1+z) \label{eqn:mbiasdefsz}\\
&=&{\rm ln}\Mbias_0+{\rm ln}\Mbias(z) \label{eqn:mbiasdefszb} \\
\siglnM^2(M,z)&=&\sigma_{0}^2 + \sum_{i=1}^{3}b_iz^i \label{eqn:msigdefsz}\\
&=&\sigma_{0}^2 +\siglnM^2(z) \label{eqn:msigdefszb}
\end{eqnarray}
We set the fiducial mass scatter to $\sigma_{0}=0.25$, and all the other 
nuisance parameters to zero.
In total, we use six nuisance parameters for the scatter and bias in mass
(${\rm ln}\Mbias_0$, $a_1$, $\sigma_{0}^2$, $b_i$).

We assume a DES-like optical cluster survey with fiducial mass threshold of
$\Mth=10^{13.5} h^{-1} \Msun$ and maximum redshift of 1.
\cite{koe07} and \cite{joh07} were able to detect clusters with mass greater 
than $10^{13.5}h^{-1} \Msun$ with a high level of purity and completeness 
using photometric data from the Sloan Digital Sky Survey (SDSS, \citep{yor00}).
The MaxBCG method used by these authors relies on red cluster galaxies occupying 
a distinct region in color space, the red sequence.
The red sequence is known to be present in clusters at least to redshift of $1$ 
(see e.g. \cite{gla00}), so that we are justified in our choice for the 
expected DES mass threshold.  
Our choice of maximum redshift is somewhat conservative since 
with the addition of the IR filters from VISTA survey, DES+VISTA will have 
accurate redshifts (for field galaxies) up to $z\sim1.5$. 
Conversely, the maximum redshift of 2 for SPT relies on the expectation that
a deeper optical follow-up may be available for SPT-detected clusters.
We show in \S \ref{sec:res} that if the cross-calibration is performed, 
the SZ clusters above $z \sim 1$ contribute very little to the cosmological constraints.

Different studies suggest a wide-range of scatter for optical observables,
ranging from a constant $\siglnM=0.5$ \citep{wu08} to a mass-dependent scatter
in the range $0.75 < \siglnM < 1.2$ \citep{bec07}.
After the submission of this paper, a couple of papers made more optimistic estimates
for the scatter.
Using weak lensing and X-ray analysis of MaxBCG selected optical clusters \cite{roz08a} 
estimated a scatter of $\sim 0.45$ between weak lensing and optical richness estimates.
In \cite{roz08b} the authors show that improved richness estimators may reduce the 
optical scatter.
As a conservative compromise, we choose a fiducial mass scatter of $\siglnM=0.5$ and allow for a cubic 
evolution in redshift and mass:
\begin{eqnarray}
{\rm ln}\Mbias(M,z)&=&{\rm ln}\Mbias_0 + a_1(1+z)\nonumber \\
&&+a_2({\rm ln}\Mobs-{\rm ln}M_{\rm pivot}) \label{eqn:mbiasdef}\\
&=&{\rm ln}\Mbias_0+{\rm ln}\Mbias(z)+{\rm ln}\Mbias(M) \nonumber
\label{eqn:mbiasdefb} \\
\siglnM^2(M,z)&=&\sigma_{0}^2 + \sum_{i=1}^{3}b_iz^i \nonumber \\
&&+ \sum_{i=1}^{3}c_i({\rm ln}\Mobs-{\rm ln}M_{\rm pivot})^i. \label{eqn:msigdef}\\
&=&\sigma_{0}^2 +\siglnM^2(M) +\siglnM^2(z) \nonumber
\label{eqn:msigdefb}
\end{eqnarray}
\noindent We set $\ln(M_{\rm pivot})=34.5$ (with $M$ in units of $h^{-1}\Msun$). 
In all, we have 10 nuisance parameters for the optical mass errors 
(${\rm ln}\Mbias_0$, $a_1$, $a_2$, $\sigma_{0}^2$, $b_i$, $c_i$).
The results we obtain are sensitive to the choice of parametrization, 
particularly the number of nuisance parameters. 
There are few, if any, constraints on the number of parameters necessary 
to realistically describe the evolution of the variance and bias with 
mass for any technique. 
If simpler parametrizations than the ones we adopt here should prove to describe 
the variations in the errors well, than cosmological constraints would improve.

 \begin{figure*}
  \begin{minipage}[t]{70mm}
    \begin{center}
      \resizebox{70mm}{!}{\includegraphics[angle=0]{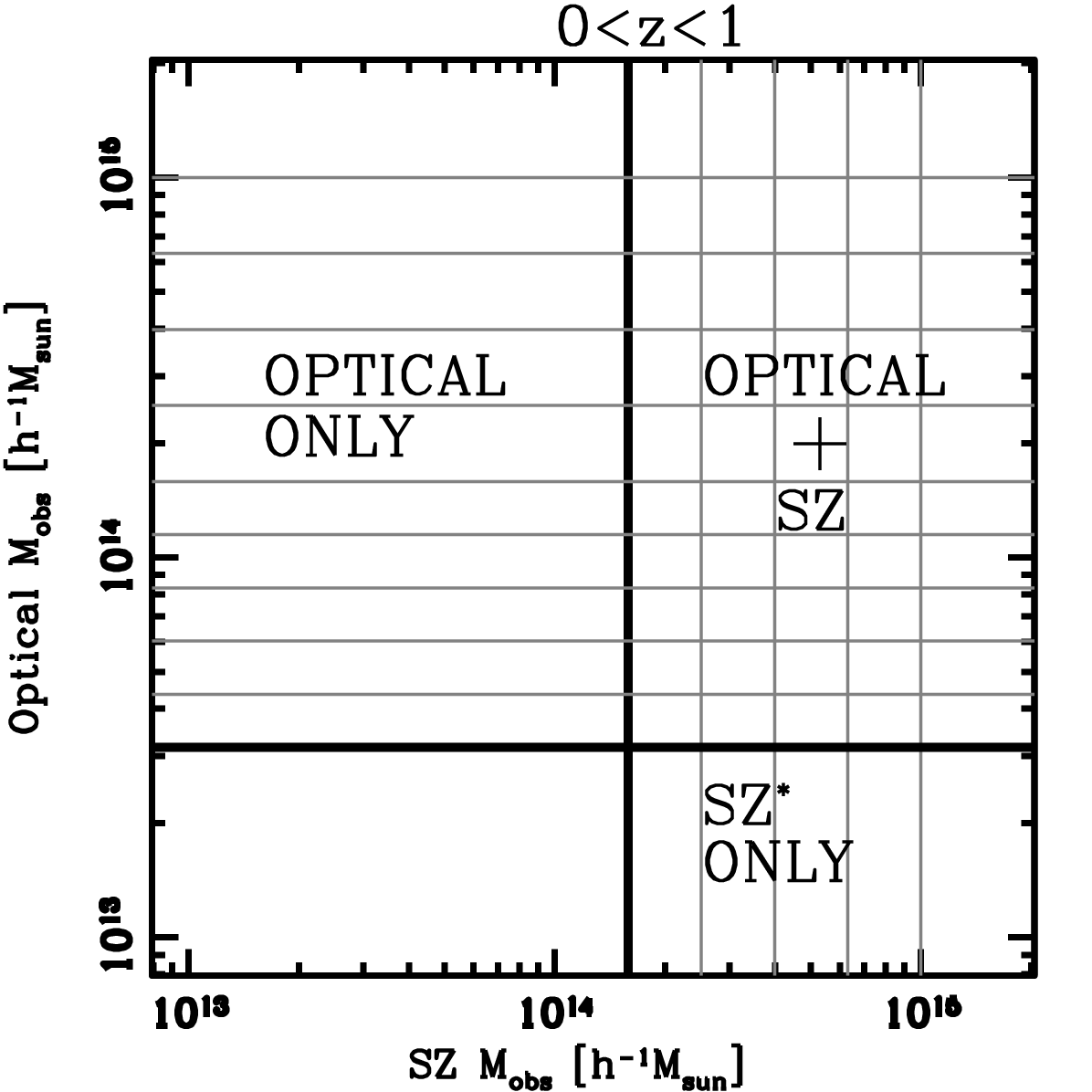}}
    \end{center}
  \end{minipage}
  \begin{minipage}[t]{70mm}
    \begin{center}
      \resizebox{70mm}{!}{\includegraphics[angle=0]{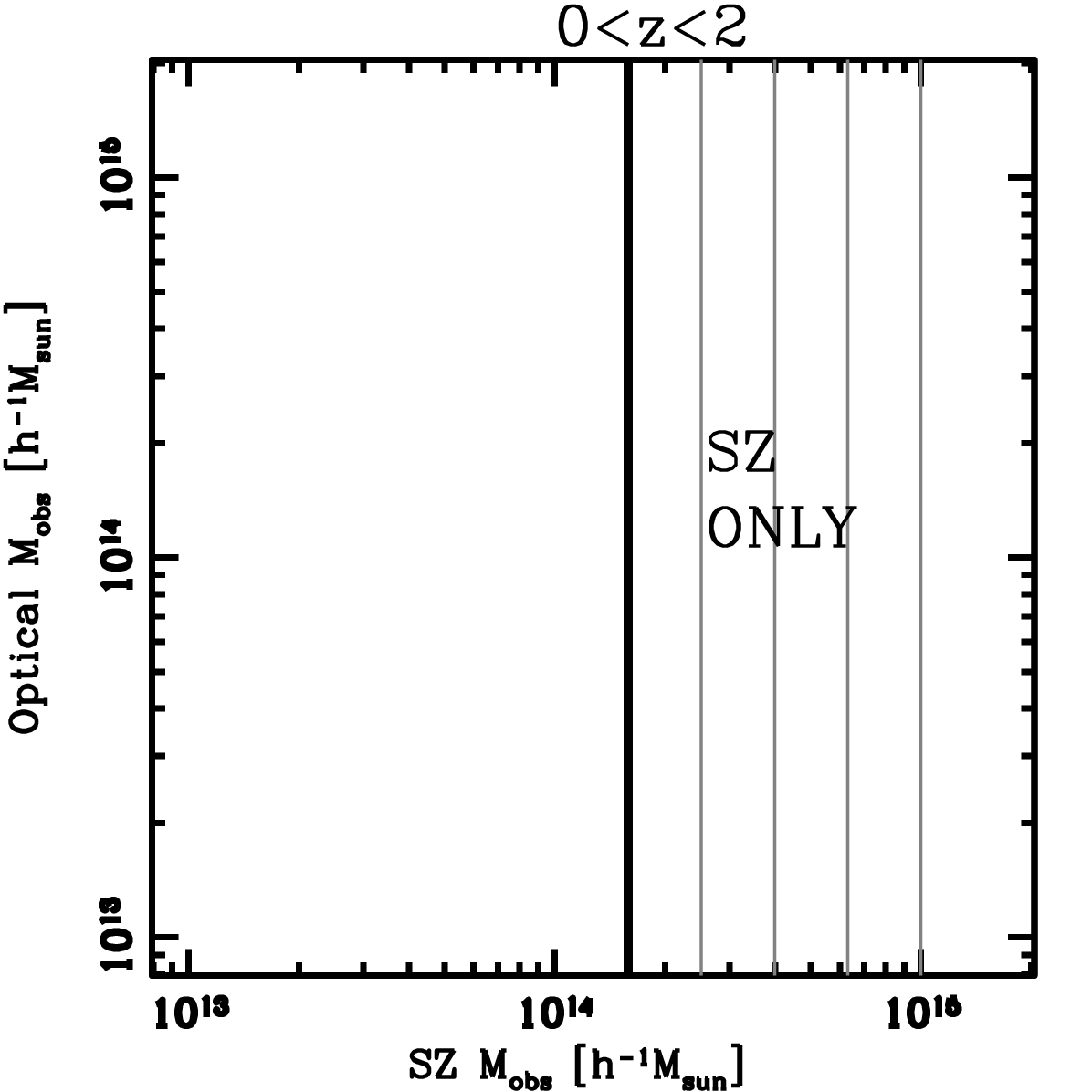}}
    \end{center}
  \end{minipage}
  \caption{ Optical-SZ mass bins in the redshift range ({\it left}) $0<z<1$ and  
    ({\it right}) $1<z<2$. The {\it black} lines indicate the mass-threshold for the SZ
    and optical surveys. The {\it gray} lines show the boundaries of the mass bins.
We do not use the SZ only region marked with the asterisk because there are very few clusters in
that region.
  }\label{fig:paramspace}
  \end{figure*}

\subsection{Redshift/observables space}\label{sec:manymass}

To calculate the SZ counts and sample variance, we use mass bins of width 
$\log(\Delta \Mobs)=0.2$ with the exception of the highest mass bin, 
which we extend to infinity.
We set the width of our redshift bins to $\Delta \zphot =0.1$.
These bin sizes imply 5 bins of mass and 20 redshift bins for the SZ clusters.
For the fiducial optical parameters, we divide the mass range 
$10^{13.5}\leq \Mobs_{\rm opt} \leq 10^{14.2} h^{-1} \Msun$ into 5 bins and use the 
same mass binning as the SZ for $\Mobs_{\rm opt} > 10^{14.2} h^{-1} \Msun$, with a total of 10 mass bins 
and 10 redshift bins.

If the clusters detected by the optical and SZ surveys are in different parts of the sky,
then the samples are independent.
To estimate the joint constraints from both surveys one simply applies the single 
mass-observable analysis described in the previous section to each of the 
samples and sums the Fisher matrices.

If the clusters are all in the same part of the sky, then the samples are not independent.
In addition, some regions of redshift/observable space contain clusters detected by 
both methods or only one. 
Our cross-calibration approach calculates the mean counts and clustering 
at all bins shown in Fig. \ref{fig:paramspace}.
From Fig. \ref{fig:paramspace} one can see that the observables parameter 
space is composed of four parts.
One is defined as the set of clusters for which 
$10^{13.5} <\Mobs_{\rm opt} < 10^{14.2} h^{-1} \Msun$,  
$\Mobs_{\rm sz} < 10^{14.2} h^{-1} \Msun$, and $0<z<1$.
Only optical clusters are detected in this region.
We divide that interval of mass into 5 equally spaced bins and use 
$P(\Mobs_{\rm opt}|M)$ to estimate the counts in that region.
The second region is defined as the clusters for which 
$\Mobs_{\rm sz} > 10^{14.2} h^{-1} \Msun$, 
$\Mobs_{\rm opt} > 10^{13.5} h^{-1} \Msun$ and $0<z<1$. 
The mass bins are simply the outer product
of the optical and SZ vectors of bins of observables in that range.  
It is comprised of $5\times 10$ mass bins and $10$ redshift bins.
Here we use $P(\Mobs_{\rm opt},\Mobs_{\rm sz}|M)$ to estimate the counts. 
The third region is defined by $\Mobs_{\rm sz} > 10^{14.2} h^{-1} \Msun$, 
$\Mobs_{\rm opt} < 10^{13.5} h^{-1} \Msun$ and $0<z<1$.
Because there are almost no clusters detected in this region, 
we do not include it in our analysis.
The fourth region is defined by $\Mobs_{\rm sz} > 10^{14.2} h^{-1} \Msun$ and $1<z<2$.
Since only SZ clusters can be found in this region we estimate the counts using 
$P(\Mobs_{\rm sz}|M)$.
The counts from the three regions we use are organized into a single vector 
of counts, and the corresponding covariance of the data (defined in 
Eq. \ref{eqn:covdat}) is given by a single matrix.

Fig. \ref{fig:counts.z} hints that our choice of binning results 
in a large number of bins with mean counts substantially below unity.
Such small number of clusters per bin brings about two concerns.
The first is that in a real survey one would not be able to accurately 
estimate the mean of such bins.
While this is true, our goal in this paper is to examine how much information 
is in the counts, which we can only be certain of extracting using 
a large number of bins. 
Our choice of binning does not yield overly optimistic results since the shot noise increases
as the counts per bin become smaller.
The bins with very few objects therefore do not contribute significantly to the Fisher matrix.
We tested this using a total of 32 bins instead of 50 (in the region of overlap of the surveys) 
and found only negligible differences in the resulting dark energy constraints.
When performing this analysis on real datasets, one would be advised to adopt a different 
binning strategy, perhaps using tree-structure algorithms to optimally subdivide the data, or
hierarchical Bayesian classification algorithms, especially if more than two observables
are used.

The second concern is that with few objects per bin the Gaussian approximation assumed when
we defined Eq. (\ref{eqn:fishmat}) - see \cite{lim04} for a derivation - is not valid.
To test the impact of the Gaussian assumption, we performed the single-observable self-calibration
analysis for the SZ survey using 5, 10, and 40 mass bins.
The results are virtually identical if 5 or 10 bins are used, but degrade by a few percent for 40 bins.
We did not investigate whether the degradation was a result of the breakdown of the Gaussian assumption
or simply due to numerical noise.
The important point is that excessive binning does not yield unrealistic improvements in the constraints.

\section{Results }\label{sec:res}

Unless stated otherwise, all results shown assume no priors on the nuisance parameters.

\subsection{Results for a single observable}\label{sec:res.1mobs}

\begin{figure*}
  \begin{minipage}[t]{70mm}
    \begin{center}
      \resizebox{70mm}{!}{\includegraphics[angle=0]{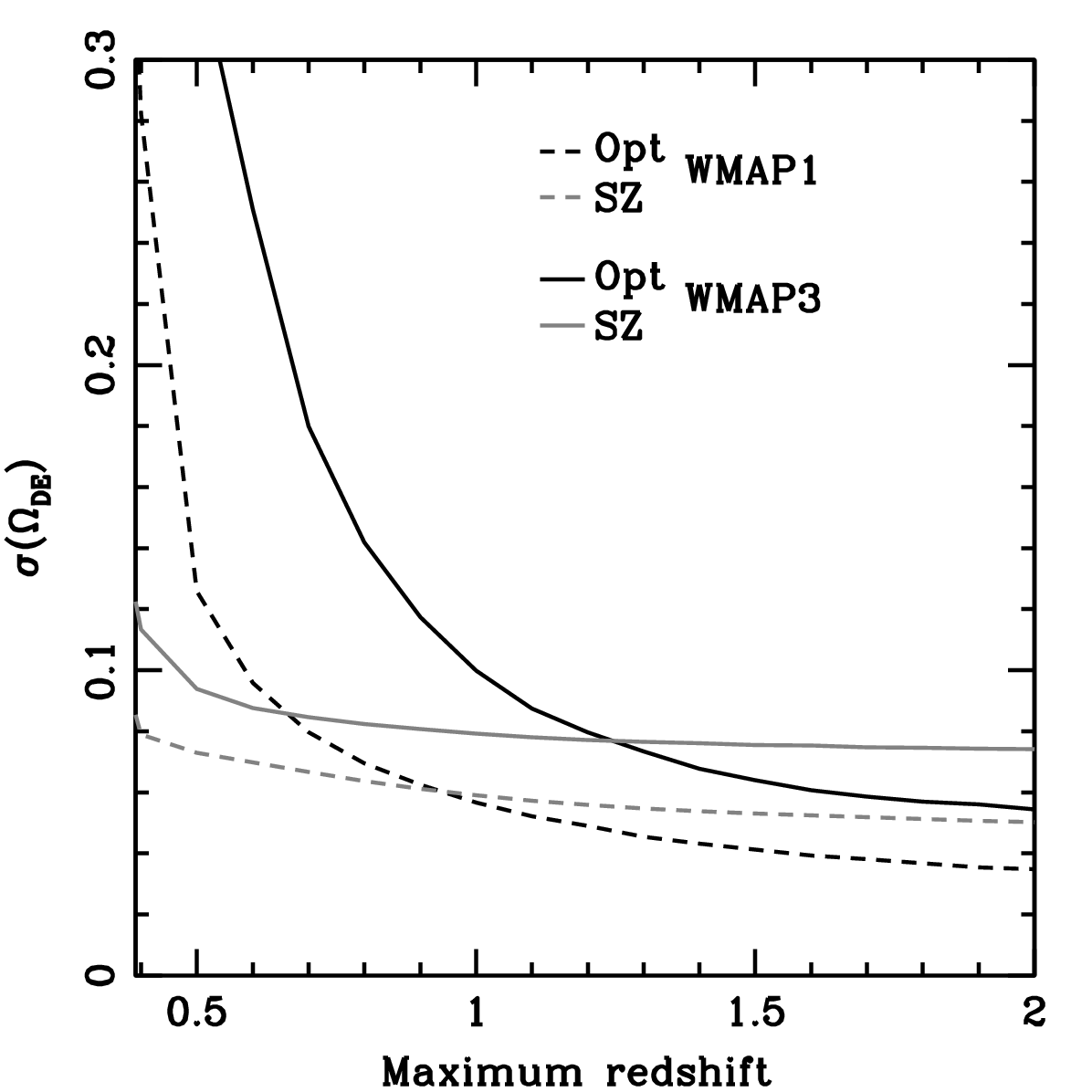}}
    \end{center}
  \end{minipage}
  \begin{minipage}[t]{70mm}
    \begin{center}
      \resizebox{70mm}{!}{\includegraphics[angle=0]{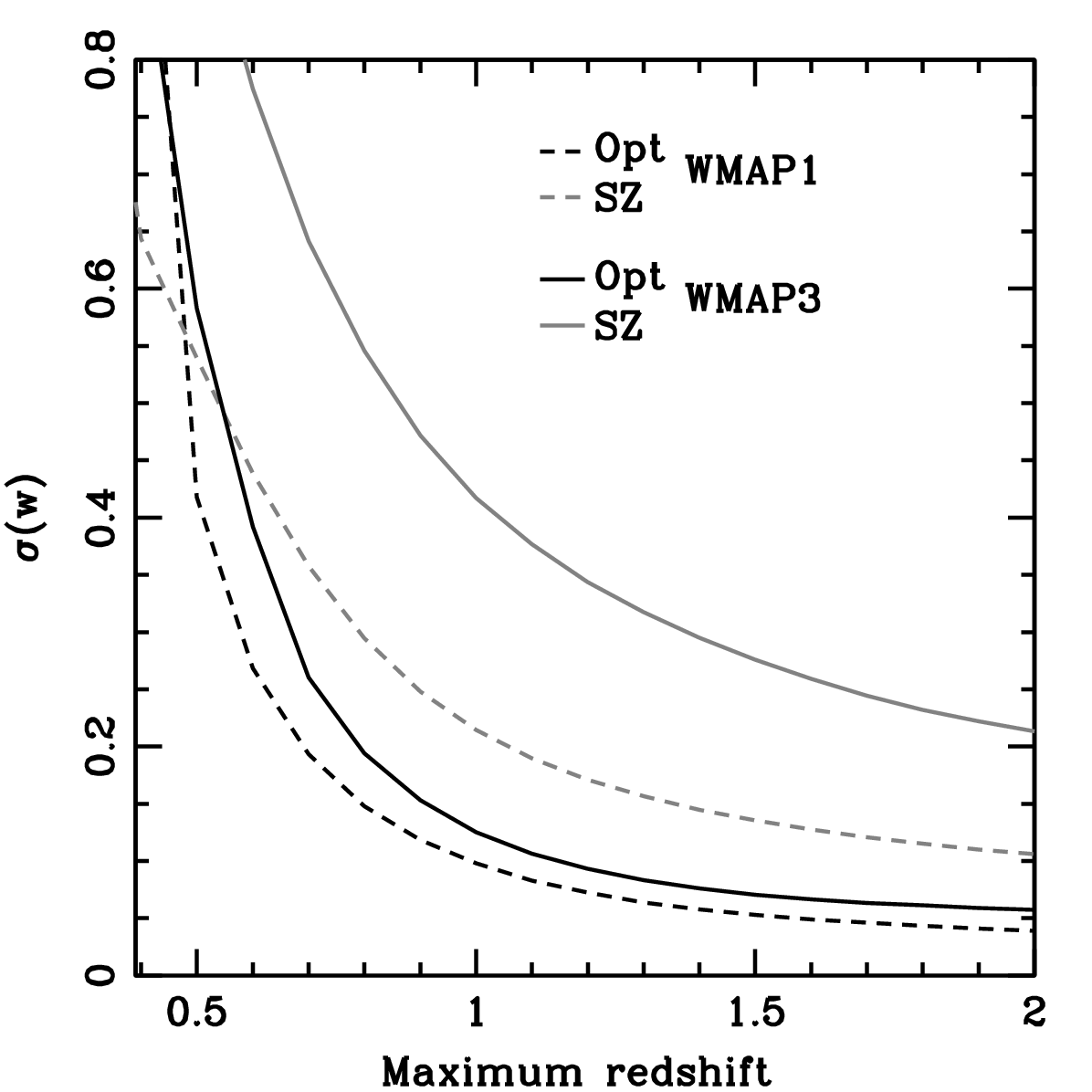}}
    \end{center}
  \end{minipage}
  \caption{Constraints on ({\it left}) $\DE$ and ({\it right}) $w$ {\it versus}
the maximum redshift of the survey for the fiducial optical and SZ surveys in WMAP1 and WMAP3
cosmologies.
  }\label{fig:sig.z}
  \end{figure*}

\begin{figure*}
  \begin{minipage}[t]{70mm}
    \begin{center}
      \resizebox{70mm}{!}{\includegraphics[angle=0]{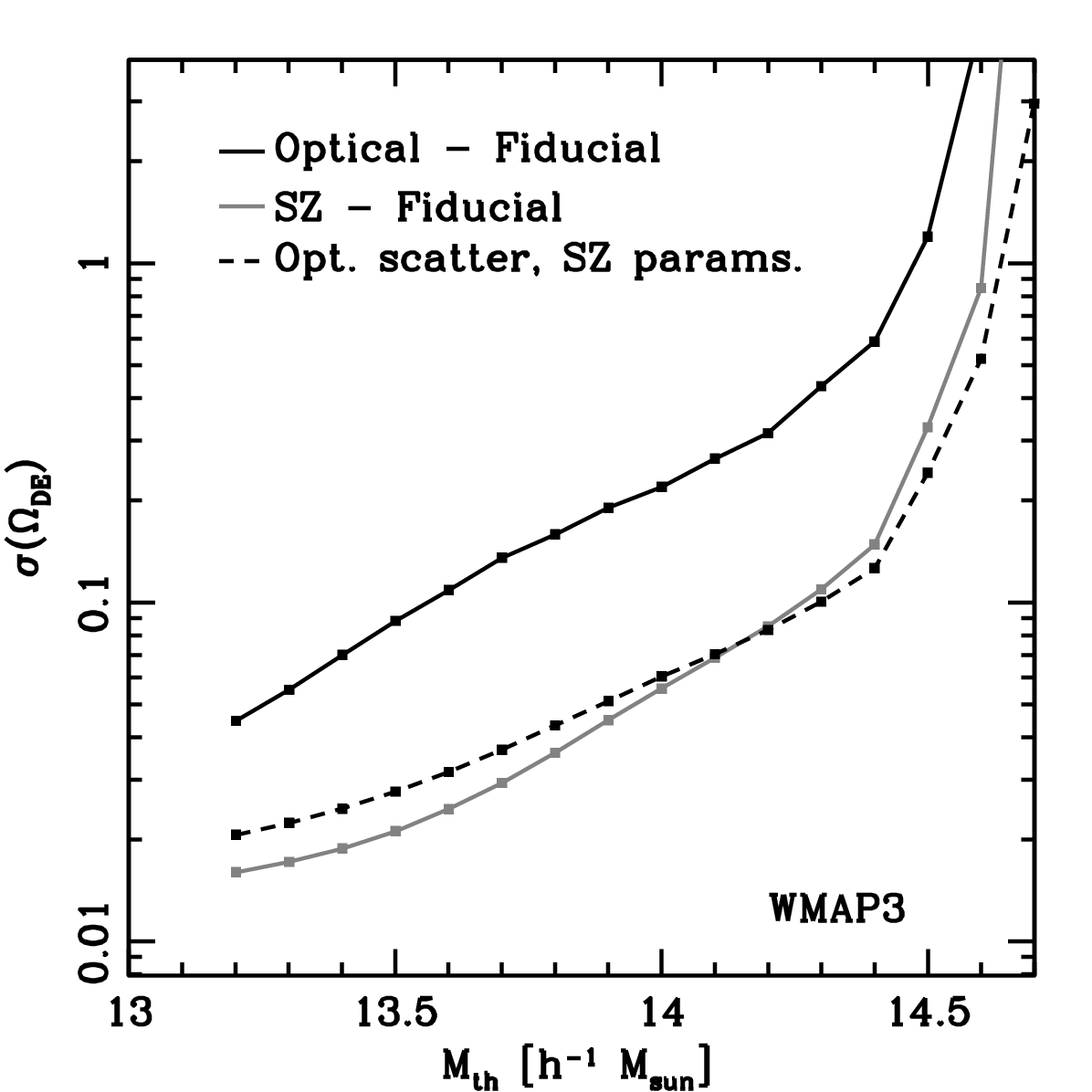}}
    \end{center}
  \end{minipage}
  \begin{minipage}[t]{70mm}
    \begin{center}
      \resizebox{70mm}{!}{\includegraphics[angle=0]{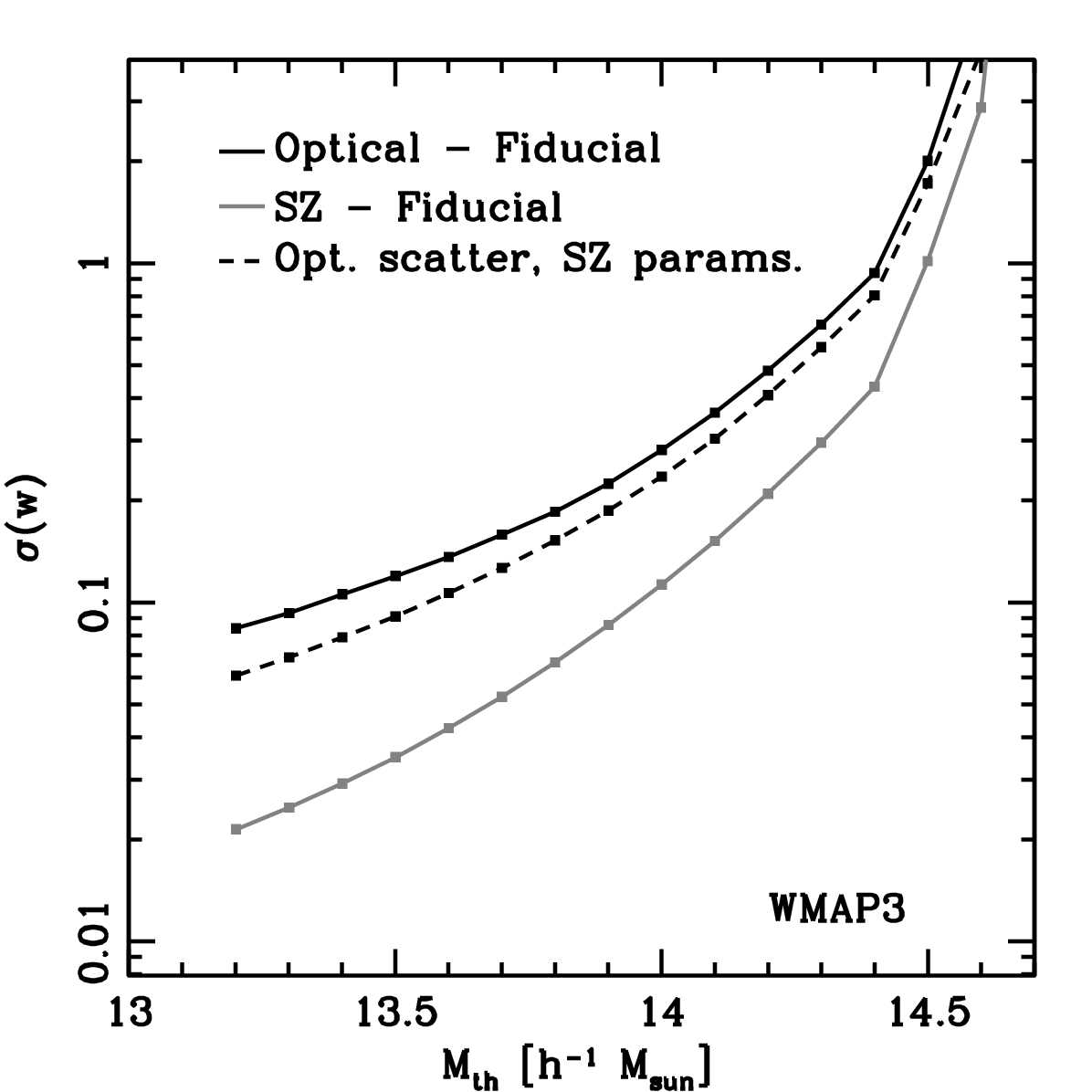}}
    \end{center}
  \end{minipage}
  \caption{Constraints on ({\it left}) $\DE$ and ({\it right}) $w$ {\it versus} 
the mass threshold of the survey in a WMAP3 cosmology.
The number of mass bins used in the calculation is different for each $\Mth$.
At the lowest threshold $\Mth=10^{13.2} h^{-1} \Msun$ and 16 bins of $\Mobs$ are used. 
We increase $\Mth$ in steps of $\Delta \ln \Mobs =0.1$ and decrease the number 
of mass bins by one at every step up to $\Mth=10^{14.7} h^{-1} \Msun$.
The {\it solid black} and {\it solid gray} lines are the marginalized 
constraints for the fiducial optical and SZ parametrizations. 
For the {\it dashed black} line we assume no mass dependence in the optical mass scatter, i.e. 
it uses the exact same parametrization as the fiducial SZ survey, except that $\sigma_0=0.5$
and the maximum redshift is 1.
  }\label{fig:sig.mth}
  \end{figure*}

First, we present results for a single observable.
Figure \ref{fig:sig.z} shows the dependence of the constraints 
on $\Omega_{\rm DE}$ ({\it left}) and $w$ ({\it right}) on the 
maximum redshift of the survey ($z_{\rm max}$).
The {\it dashed} and {\it solid black} lines are for the fiducial 
optical mass threshold, scatter and bias in WMAP1 and WMAP3 
cosmologies, respectively.
The {\it dashed} and {\it solid gray} lines are the corresponding results 
assuming the fiducial SZ survey.
The rate of improvement in the $\DE$ constraints with $z_{\rm max}$ 
decreases sharply after $z\sim 0.5$ for all cases except the optical 
results in WMAP3, where the break happens 
around $z\sim1$.
The constraints on $w$ show a more pronounced redshift dependence 
for both optical and SZ.
In a WMAP3 cosmology, varying $z_{\rm max}$ from 1 to 2 results in 
$\sigma(w)$ decreasing by a factor of $\sim 2.5$ for the optical
and $\sim 2.1$ for the SZ.
The intersection of the dashed lines in both plots, or of the solid lines
in the {\it left} plot mark the redshifts below which
the optical survey yields tighter constraints than the SZ survey.
At this point, Poisson noise in the counts is the dominant component of the error budget.
The increase in counts due to the larger scatter of the optical observable compensates
for the loss of information due to increased scatter. 

Figure \ref{fig:sig.mth} shows ({\it left}) $\sigma(\DE)$ and 
({\it right}) $\sigma(w)$ {\it versus} 
the mass threshold of the survey in a WMAP3 cosmology.
The number of mass bins used in the calculation is different for each $\Mth$.
At the lowest threshold $\Mth=10^{13.2} h^{-1} \Msun$ and there 16 bins of $\Mobs$. 
We increase $\Mth$ in steps of $\Delta \ln \Mobs =0.1$ and decrease the number 
of mass bins by one at every step up to $\Mth=10^{14.7} h^{-1} \Msun$.
The {\it solid black} and {\it solid gray} lines show the marginalized constraints for the 
fiducial optical and SZ parametrizations. 
For the {\it dashed black} line we assume no mass dependence in the optical mass scatter, i.e. 
we use the same parametrization as the fiducial SZ survey, except 
that $\sigma_0=0.5$, and the maximum redshift is 1.
The fact that the {\it dashed black} line drops below the {\it gray} line in 
the {\it left} plot is another illustration of the point made in \S \ref{sec:techniques} 
of larger scatter resulting in better cosmological constraints, despite the lower 
redshift range of the optical survey and no priors on the scatter.
Allowing for mass dependence of $\ln \Mbias_{\rm opt}$ and $\varopt$ not only 
degrades $\sigma(\DE)$ but also increases the sensitivity of the constraints to $\Mth$.
The constraints on $w$ are much less affected, because of the low maximum redshift of 
the optical survey.

\subsection{Results for two observables}\label{sec:res.2mobs}

\begin{figure*}
  \begin{minipage}[t]{70mm}
    \begin{center}
      \resizebox{70mm}{!}{\includegraphics[angle=0]{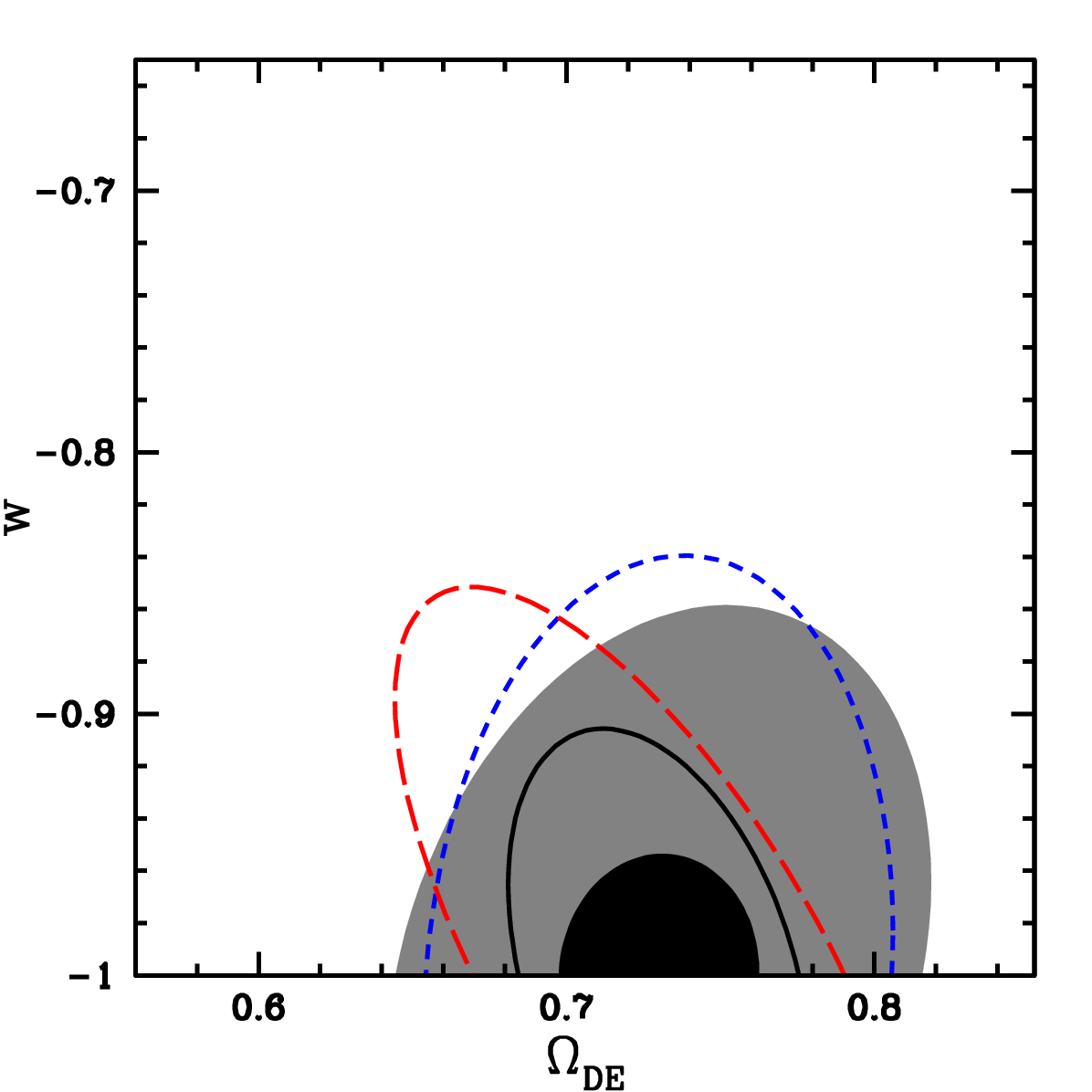}}
    \end{center}
  \end{minipage}
  \begin{minipage}[t]{70mm}
    \begin{center}
      \resizebox{70mm}{!}{\includegraphics[angle=0]{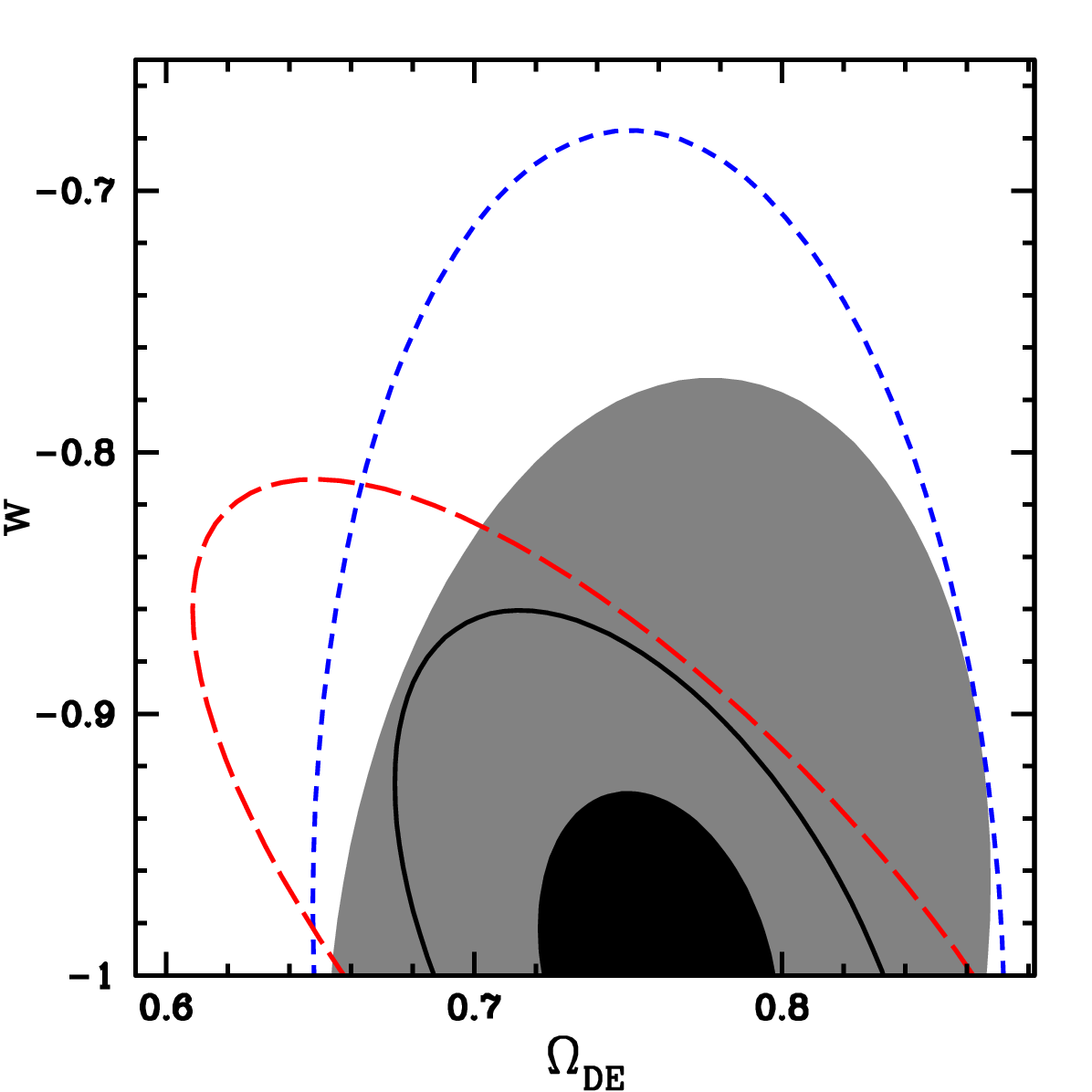}}
    \end{center}
  \end{minipage}
  \caption{$68\%$ confidence regions in the $\DE-w$ plane in ({\it left}) WMAP1 
and ({\it right}) WMAP3 cosmologies.
The constraints from cross-calibration using only clusters detected simultaneously 
in optical and SZ (i.e. {\it partial cross-calibration} - with selection
$\Mobs_{\rm sz}>10^{14.2} h^{-1} \Msun$, $\Mobs_{\rm opt}>10^{13.5} h^{-1} \Msun$  and 
$0<z<1$) are represented by the {\it filled gray} ellipses.
The cross-calibration using all clusters ({i.e. \it full cross-calibration}) yields the 
{\it filled black} ellipses.
For comparison, the {\it long dashed red} lines show constraints for the fiducial optical survey,
and the {\it short dashed blue} lines show constraints for the fiducial SZ survey.
Treating the optical and SZ surveys as independent and adding their Fisher matrices yields
the {\it solid black} lines.
}\label{fig:ellipse.optsz}
\end{figure*}

\begin{figure}
    \begin{center}
      \resizebox{70mm}{!}{\includegraphics[angle=0]{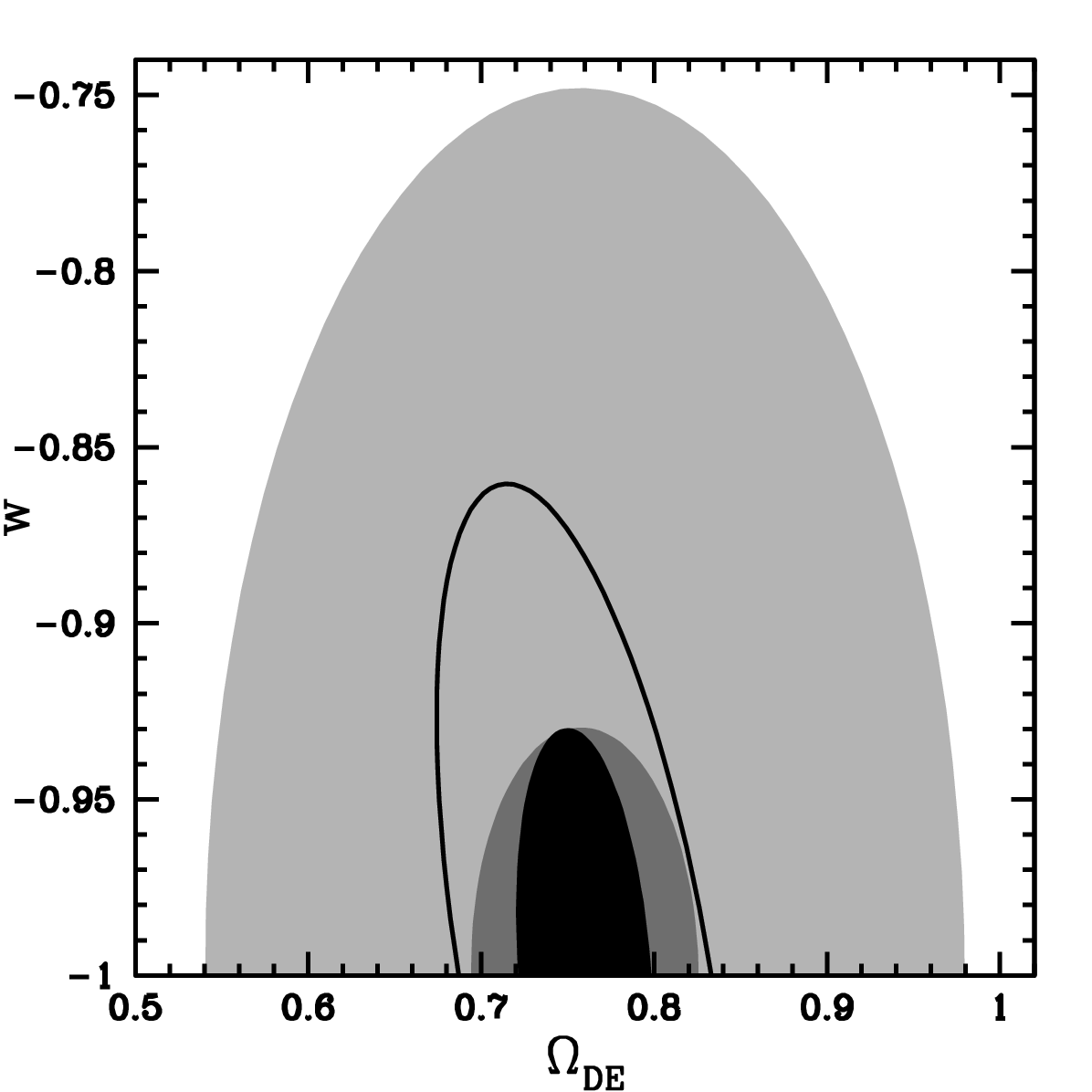}}
    \end{center}
\caption{The {\it filled light gray} ellipse shows the constraints from summing the SZ and optical fisher matrices without
clustering.
The {\it solid black} line  indicates the corresponding constraints when clustering is added. 
The {\it filled dark gray} and {\it filled black} ellipses show the full cross-calibration constraints
without and with clustering, respectively.
}\label{fig:ellipse.optsz.noclust}
\end{figure}

\begin{table}
\caption{Marginalized constraints on cosmological parameters}
\begin{center}
\leavevmode
\begin{tabular}{ l c c c c }\hline \hline
\multicolumn{1}{c}{} & \multicolumn{2}{c}{WMAP1} & \multicolumn{2}{c}{WMAP3} \\
\hline
Survey & $\sigma(\DE)$ & $\sigma(w)$ & $\sigma(\DE)$ & $\sigma(w)$ \\
\hline
Intersection &0.058&0.093&0.070& 0.15\\
Optical  &0.057&0.098 &0.10&0.13\\
SZ &0.050 & 0.11 & 0.074 & 0.21\\
Optical + SZ  &0.032 & 0.062 & 0.057 & 0.092\\
Cross-Cal. (Full)$^c$&  0.021 &  0.030 & 0.025 &  0.045 \\
Cross-Cal. ($z^{\rm SZ}_{max}<1.1$)$^c$ & 0.022 &  0.032 & 0.026 &  0.047\\ 
\hline \hline
\label{tbl:constraints}
\end{tabular}
\end{center} 
\begin{flushleft}
 $^c$ Fixed $\rho=0$\\
\end{flushleft}
\end{table}

Figure \ref{fig:ellipse.optsz} shows the $68\%$ confidence regions for $\Omega_{\rm DE}$ 
and $w$ in ({\it left}) WMAP1 and ({\it right}) WMAP3 cosmologies assuming no priors 
in the nuisance parameters and no correlation between the observables (i.e. $\rho=0$, fixed).
Comparing both plots, we see that the low fiducial number of clusters in the WMAP3 
cosmology implies weaker cosmological constraints.
More interestingly, in a cosmology with fewer clusters the lower mass threshold of 
the optical technique makes it more constraining than the fiducial SZ 
even without any priors on the bias or scatter.
The marginalized constraints are summarized in Table \ref{tbl:constraints}.

Performing the cross-calibration using only clusters detected by both methods (hereafter 
{\it partial cross-calibration} - represented in the plots by the {\it filled gray} ellipses) 
does not yield very good constraints.
The partial cross-calibration is slightly more useful in a WMAP3 cosmology, because 
there are few clusters above $z=1$, so that not using that region of parameter space does not
cause much degradation. 
Constraints using the cross-calibration with all clusters available (hereafter 
{\it full cross-calibration} - {\it filled black} ellipses) 
yields much better constraints than the partial cross-calibration.
In fact, constraints on ${\DE}$ and ${w}$ from the full cross-calibration 
are a factor $\sim 2$ better than constraints derived by simply adding the Fisher matrices of 
the optical and SZ techniques (the {\it solid black line}).

We demonstrate the importance of clustering in a WMAP3 cosmology to self- and cross-calibration in 
Fig. \ref{fig:ellipse.optsz.noclust}.
Comparing the {\it filled light gray} ellipse with the {\it solid black} line, we see that clustering
information tightens constraints on both $\DE$ and $w$ significantly if we only sum the  
optical and SZ Fisher matrices.
But comparing the {\it filled dark gray} ellipse with the {\it filled black} ellipse we
see that clustering does not add as much information to the full cross-calibration.
Constraints on $w$ are unchanged, and $\DE$ constraints improve by a factor of $\sim 1.7$.

\begin{figure}
  \begin{center}
      \resizebox{70mm}{!}{\includegraphics[angle=0]{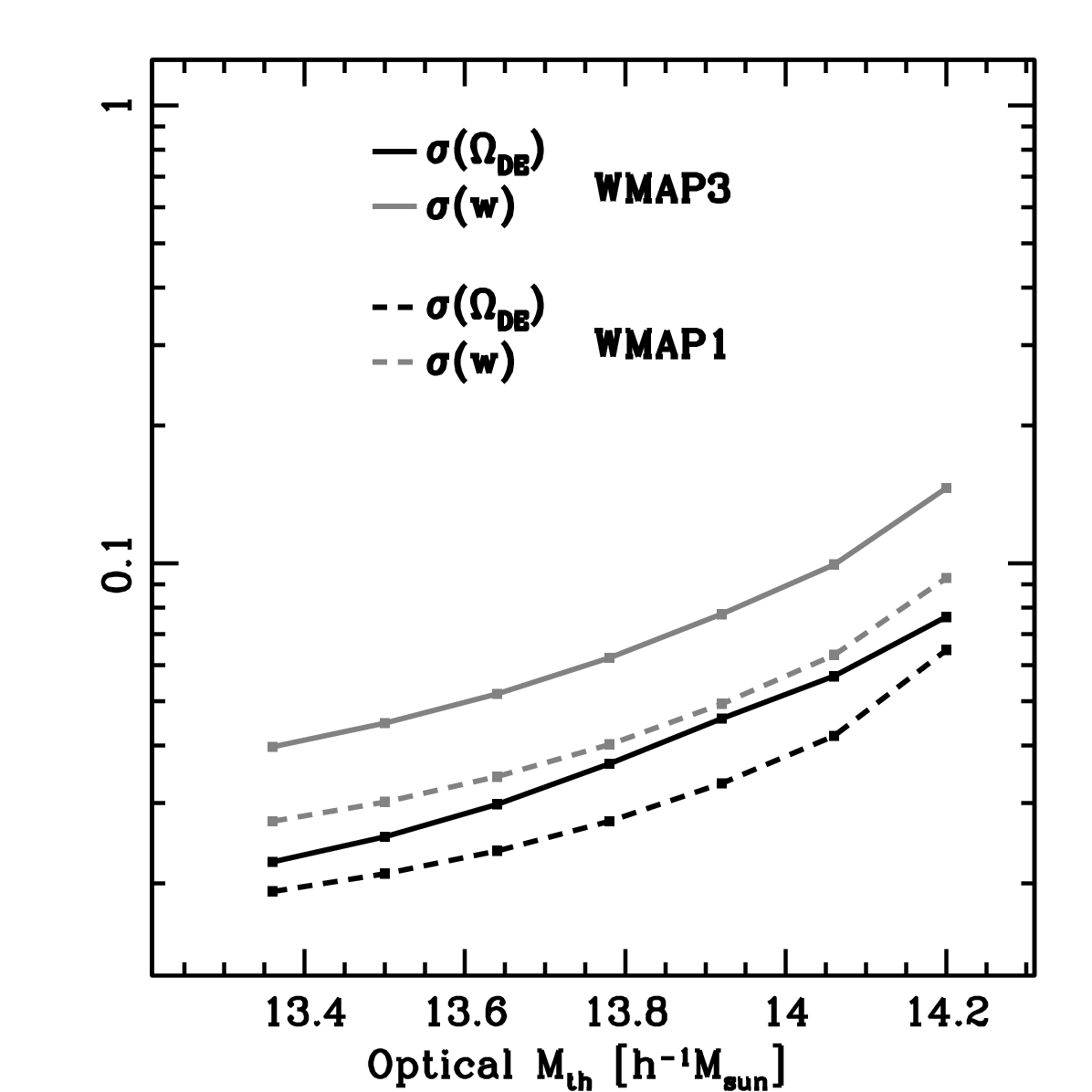}}
    \end{center}
  \caption{$\sigma(\Omega_{\rm DE})$ and $\sigma(w)$
for the full cross-calibration as a function of the optical mass threshold, 
$M_{\rm th}^{\rm opt}$ in both WMAP1 and WMAP3 cosmologies with correlation $\rho$ 
fixed at zero.
The dots indicate boundaries of the mass bins for $\Mobs_{\rm opt}<10^{14.2} h^{-1}\Msun$.
Above $10^{14.2}$ we use the same bins as for $\Mobs_{\rm sz}$.
}\label{fig:varmth.optsz}
\end{figure}

\begin{figure}
  \begin{center}
      \resizebox{70mm}{!}{\includegraphics[angle=0]{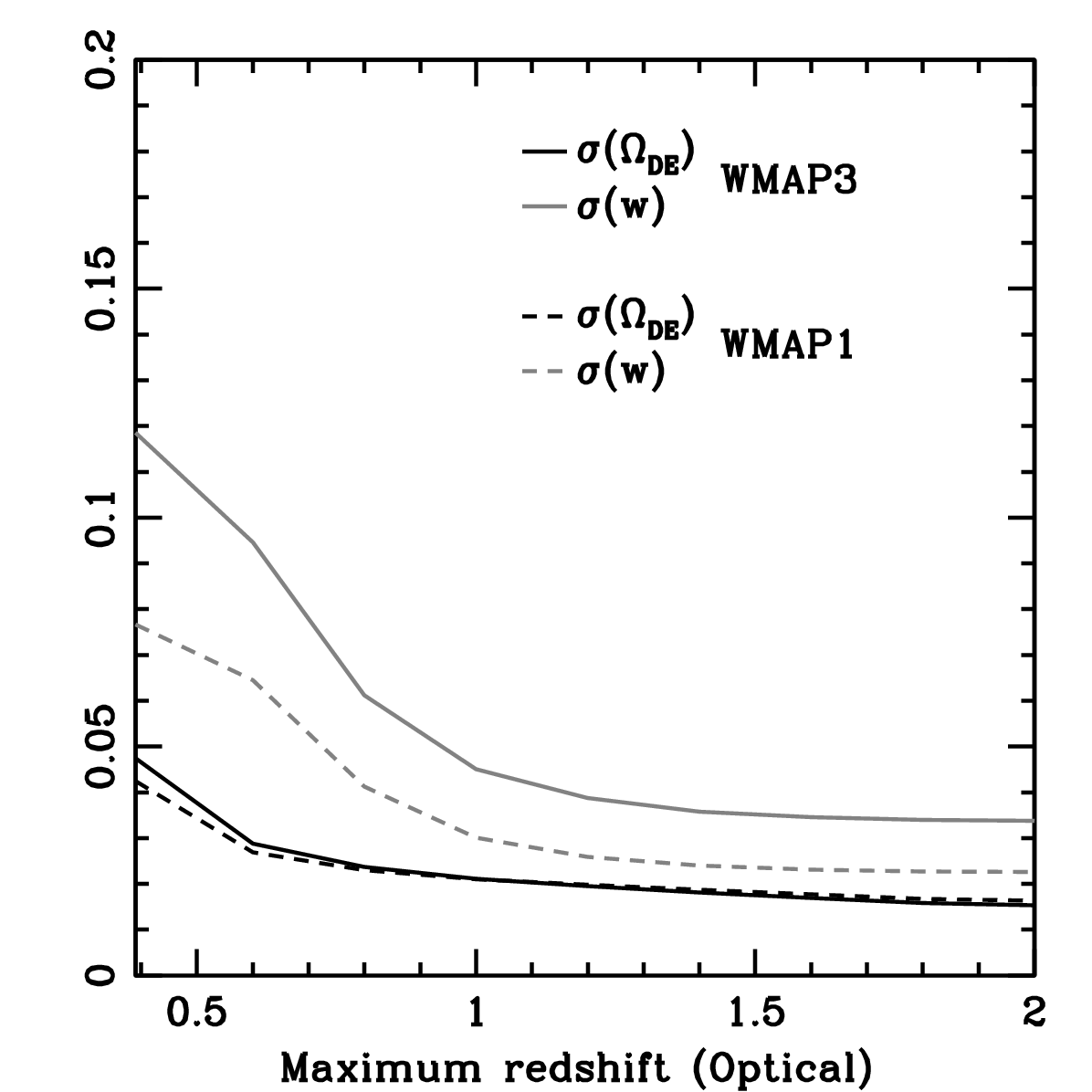}}
    \end{center}
  \caption{$\sigma(\Omega_{\rm DE})$ and $\sigma(w)$
for the full cross-calibration as a function of the maximum redshift of the optical survey, 
in WMAP1 and WMAP3 cosmologies with correlation $\rho$ fixed at zero.
}\label{fig:varz.optsz}
\end{figure}

Figure \ref{fig:varmth.optsz} shows $\sigma(\Omega_{\rm DE})$ and $\sigma(w)$
for the full cross-calibration as a function of the optical mass threshold, 
$M_{\rm th}^{\rm opt}$, in both WMAP1 and WMAP3 cosmologies with $\rho$ fixed at zero.
The dots indicate boundaries of the mass bins for $\Mobs_{\rm opt}<10^{14.2} h^{-1}\Msun$.
Above $10^{14.2}$ we use the same bins as the $\Mobs_{\rm sz}$.
Constraints on $w$ are slightly less sensitive to $\Mobs_{\rm opt}$ than constraints
on $\Omega_{\rm DE}$.
Comparing the slopes of the curves in Figure \ref{fig:varmth.optsz} and 
Figure \ref{fig:sig.mth} we see that the full cross-calibration constraints are 
less sensitive to $\Mth$ than the self-calibrated constraints from optical or SZ alone. 
In Fig. \ref{fig:sig.mth} a change in $\Mth$ from $10^{13.5}h^{-1}\Msun$ to $10^{14.2}h^{-1}\Msun$ 
results in a degradation of $\sigma(w)$ and $\sigma(\DE)$ of $\sim 4.0$ and 
$\sim 3.6$, respectively, for optical only, and of $\sim 5.9$ and $\sim 4.0$ for
SZ only.
With the full cross-calibration, the degradation factor is only $\sim 3.0$ for $\sigma(\DE)$
and $\sim 3.3$ for $\sigma(w)$.

The full cross-calibration also reduces the sensitivity to the maximum redshift range of the 
surveys.
Figure \ref{fig:varz.optsz} shows $\sigma(\Omega_{\rm DE})$ and $\sigma(w)$
as a function of the maximum redshift of the optical survey for the full 
cross-calibration.
Comparing to Figure \ref{fig:sig.z} it is clear that the individual surveys
are much more sensitive to $z_{\rm max}$ than the full cross-calibration.
For example, if  $z_{\rm max}$ changes from 1 to 2 in a WMAP3 cosmology,
the optical-only and SZ-only constraints on $w$ improve by factors of 
$\sim 2.2$ and $\sim 2.0$, respectively.
In comparison, the same change in $z_{\rm max}$ for the optical survey
in the full cross-calibration improves $w$ constraints by only $\sim 1.3$.
Cross-calibration constraints are even less sensitive to variations in 
the maximum redshift of the SZ survey. 
For a fixed optical $z_{\rm max}=1$, reducing the SZ $z_{\rm max}$ from $2$ to $1.1$ degrades
constraints by only a few percent in both cosmologies. 
In this scenario, we find $\sigma(\DE,w)=(0.022,0.048)$ in a WMAP1 cosmology and 
$\sigma(\DE,w)=(0.027,0.073)$ in a WMAP3 cosmology.

All cross-calibration results shown heretofore assumed correlation
coefficient $\rho$ fixed at zero.
From Eq. (\ref{eqn:rhosig}) we see that $\rho=0$ implies 
$\sigma_{ab}=\sigma_{\rm opt-sz}=\infty$.
Weak lensing and X-ray mass measurements of optically-selected clusters
suggest that a more realistic guess would be 
$\sigma_{\rm opt-sz}\sim 0.3-0.7$,
from which Eq. (\ref{eqn:rhosig}) implies that $0.19<|\rho|<0.55$.
A value of $\rho>0.6$ corresponds to $\sigma_{\rm opt-sz}<0.19$.
To obtain higher correlation values, one would need $\sigma_{ab}$
to be small compared to $\sigma_{a}$ and $\sigma_{b}$.

\begin{figure*}
  \begin{center}
      \resizebox{70mm}{!}{\includegraphics[angle=0]{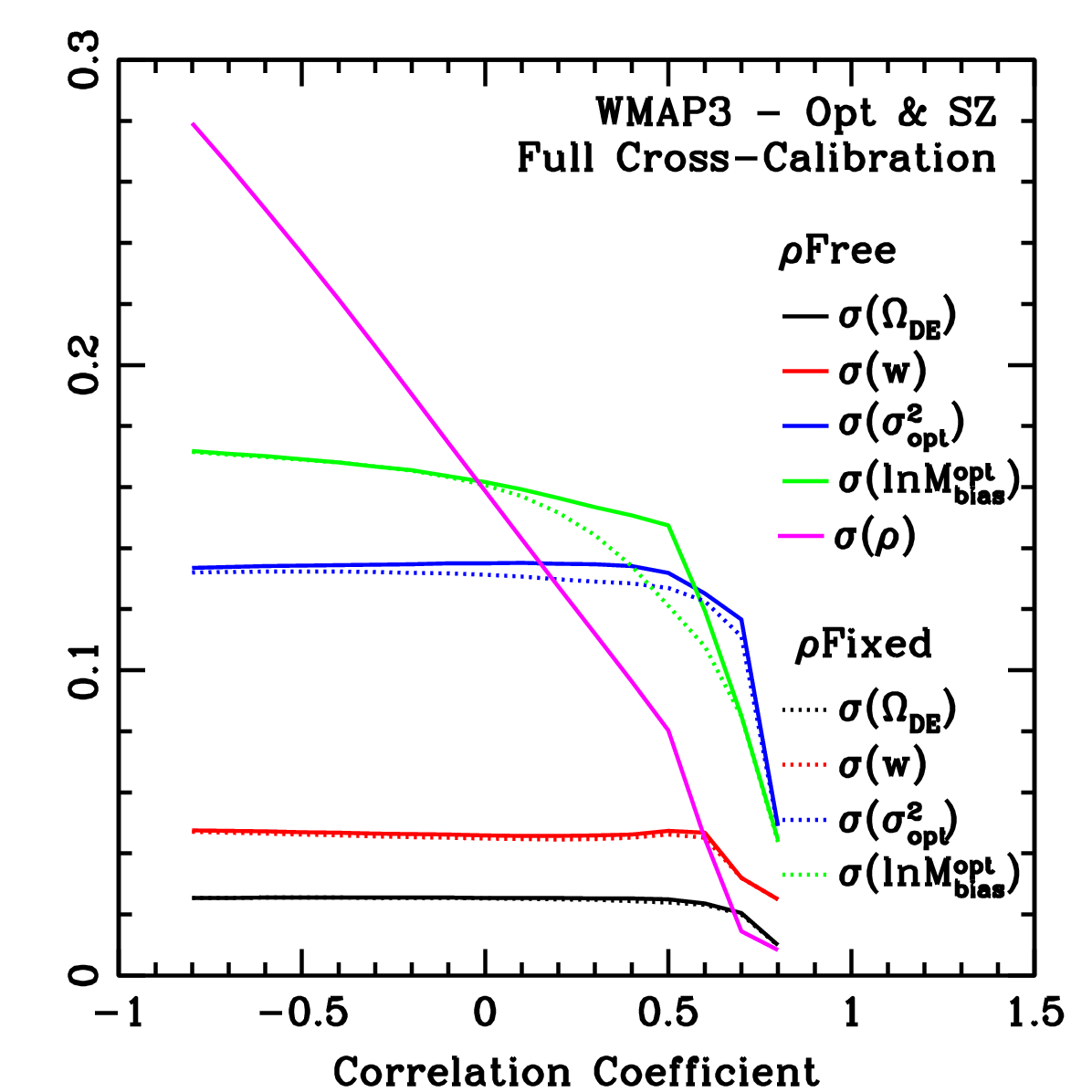}}
      \resizebox{70mm}{!}{\includegraphics[angle=0]{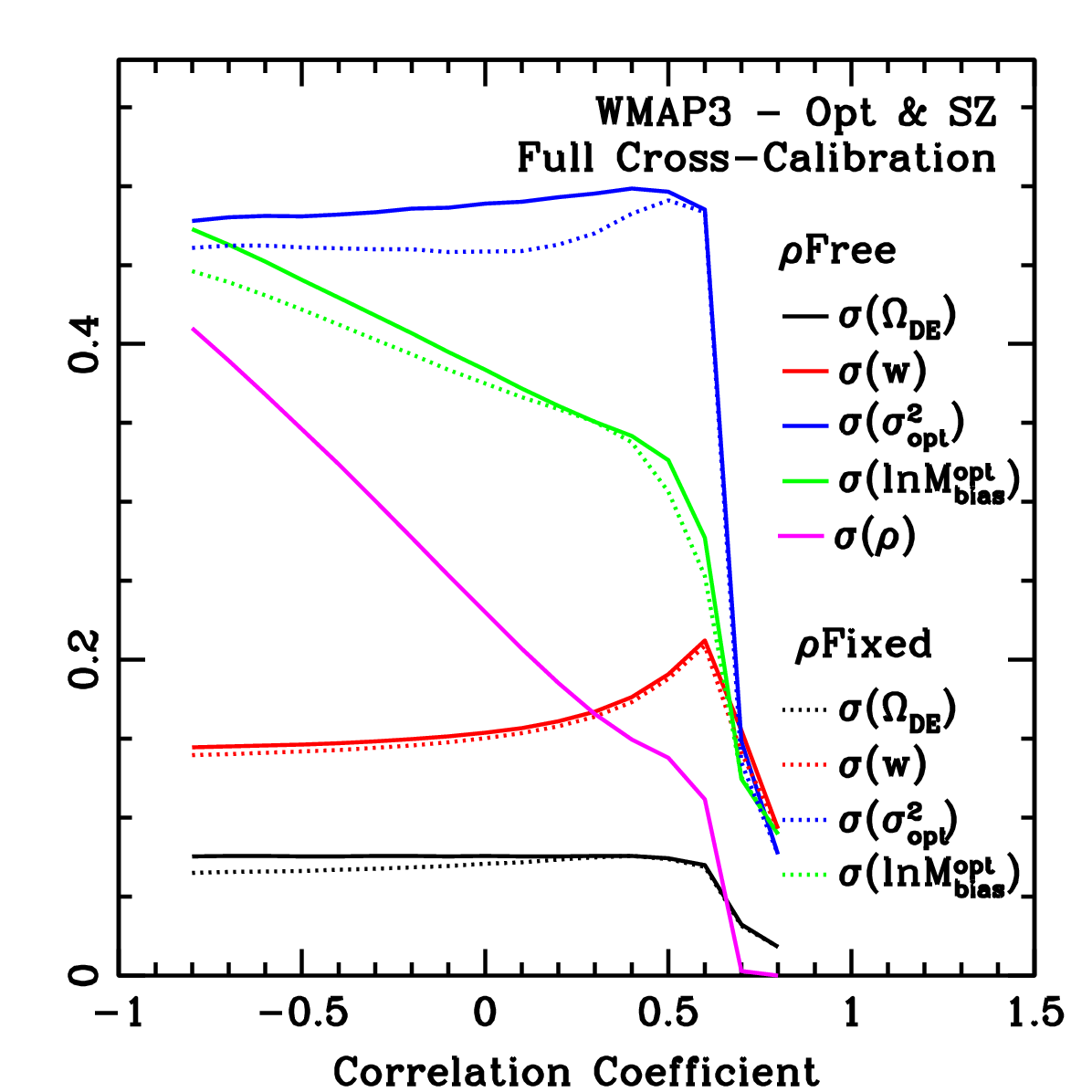}}
    \end{center}
  \caption{$1-\sigma$ constraints on dark energy and optical-mass nuisance parameters as a function
of correlation $\rho$ for ({\it left}) the full cross-calibration 
and ({\it right}) the partial cross-calibration.
For the {\it solid} lines  $\rho$ is a free parameter whereas the {\it dotted} lines are for $\rho$ fixed.
Both plots are for a WMAP3 cosmology. 
}\label{fig:rho.optsz.wmap3}
\end{figure*}

Figure (\ref{fig:rho.optsz.wmap3}) shows the dependence of the constraints
on the dark energy  and optical mass nuisance parameters on the correlation coefficient. 
From the {\it left} plot we see that the dark energy parameters are 
insensitive to the value of the correlation for $\rho<0.6$ for the full 
cross-calibration analysis. 
The very sharp drop in the uncertainties of both cosmological and nuisance
parameters is largely due to the optical and SZ surveys having different 
fiducial scatters and mass thresholds.
Given $\sigma_{\rm opt}$ and $\sigma_{\rm SZ}$, high values of the correlation
imply very low values of $\sigma_{\rm opt-sz}$, the scatter between observables.
High correlation means that the scatter in the optical is effectively that of the 
SZ survey.
From the plot we see that $\rho=0.8$, the combination of optical and SZ results 
yields constraints very similar to a survey with optical $\Mth$ but with SZ scatter 
(cf. Fig. \ref{fig:sig.mth}).

The constraints on $\rho$ improve as $\rho$ increases, though comparing 
constraints for fixed and free $\rho$, we see that dark energy constraints 
are fairly insensitive to $\sigma(\rho)$.
This means that the correlation is sufficiently well determined by the 
cross-calibration analysis without need for additional priors.

In the {\it right} plot, we see that for the cross-calibration using only
clusters detected by both methods (i.e. the partial cross-calibration) the 
constraints are more dependent on the value of the correlation and on its uncertainty.
The relation between $\rho$ and the optical bias is most pronounced.
As mentioned in the discussion following
Eq. (\ref{eqn:ewin2}), variations in the correlation change the distribution
of number counts in $\Mobs_a - \Mobs_b$ space in ways that mimic bias and scatter
in the observables.
In the full cross-calibration, the relation between $\sigma(\rho)$ and 
$\sigma(ln\Mbias_{\rm opt})$ is less pronounced because the information from
clusters detected only by optical (or SZ) helps to break the degeneracy between the
correlation and the bias.
Though not shown, the uncertainty in the bias and scatter of the SZ observable
scales very similarly to that of the corresponding optical nuisance parameters.

\begin{figure*}
\begin{center}
  \begin{minipage}[t]{70mm}
    \begin{center}
      \resizebox{70mm}{!}{\includegraphics[angle=0]{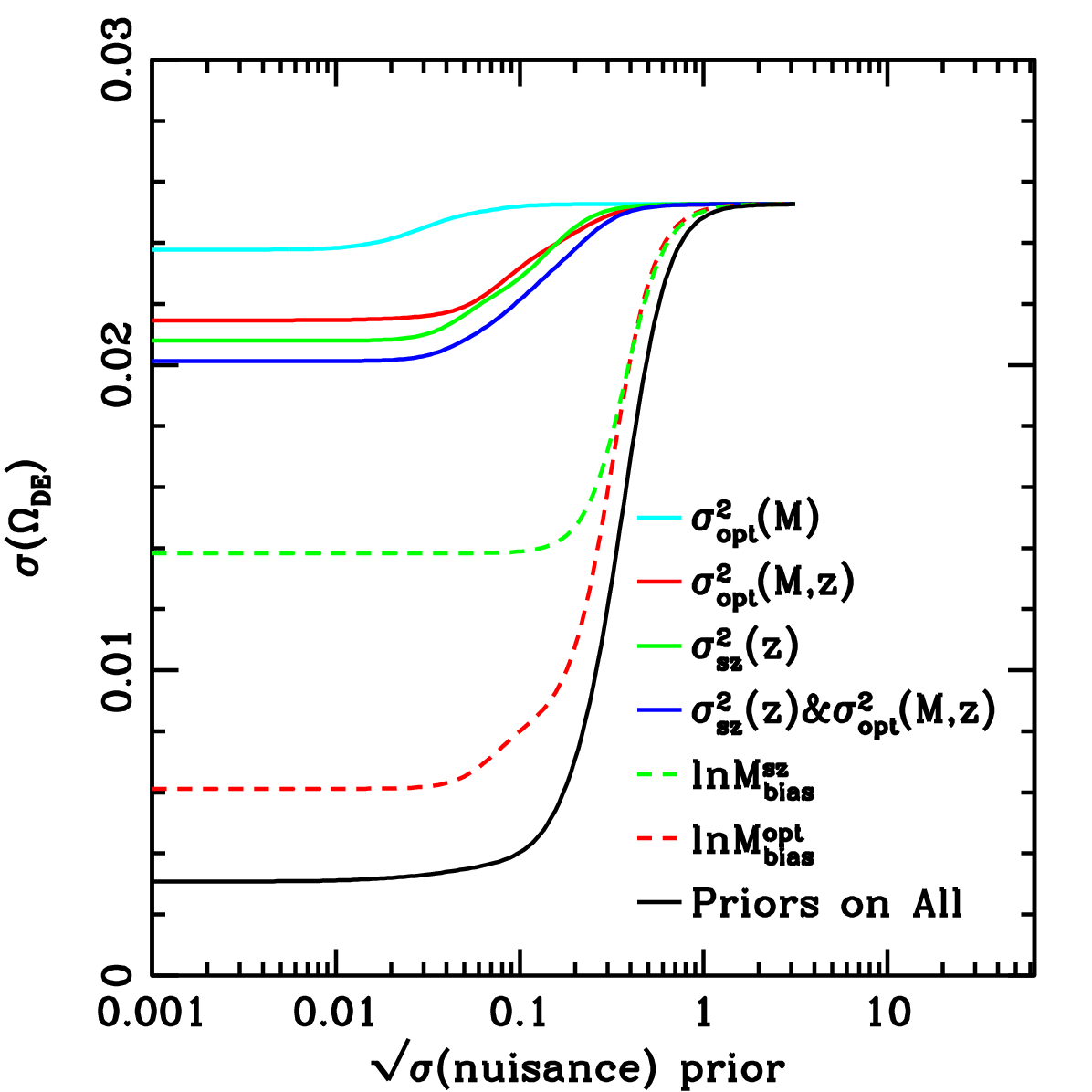}}
    \end{center}
  \end{minipage}
  \begin{minipage}[t]{70mm}
    \begin{center}
      \resizebox{70mm}{!}{\includegraphics[angle=0]{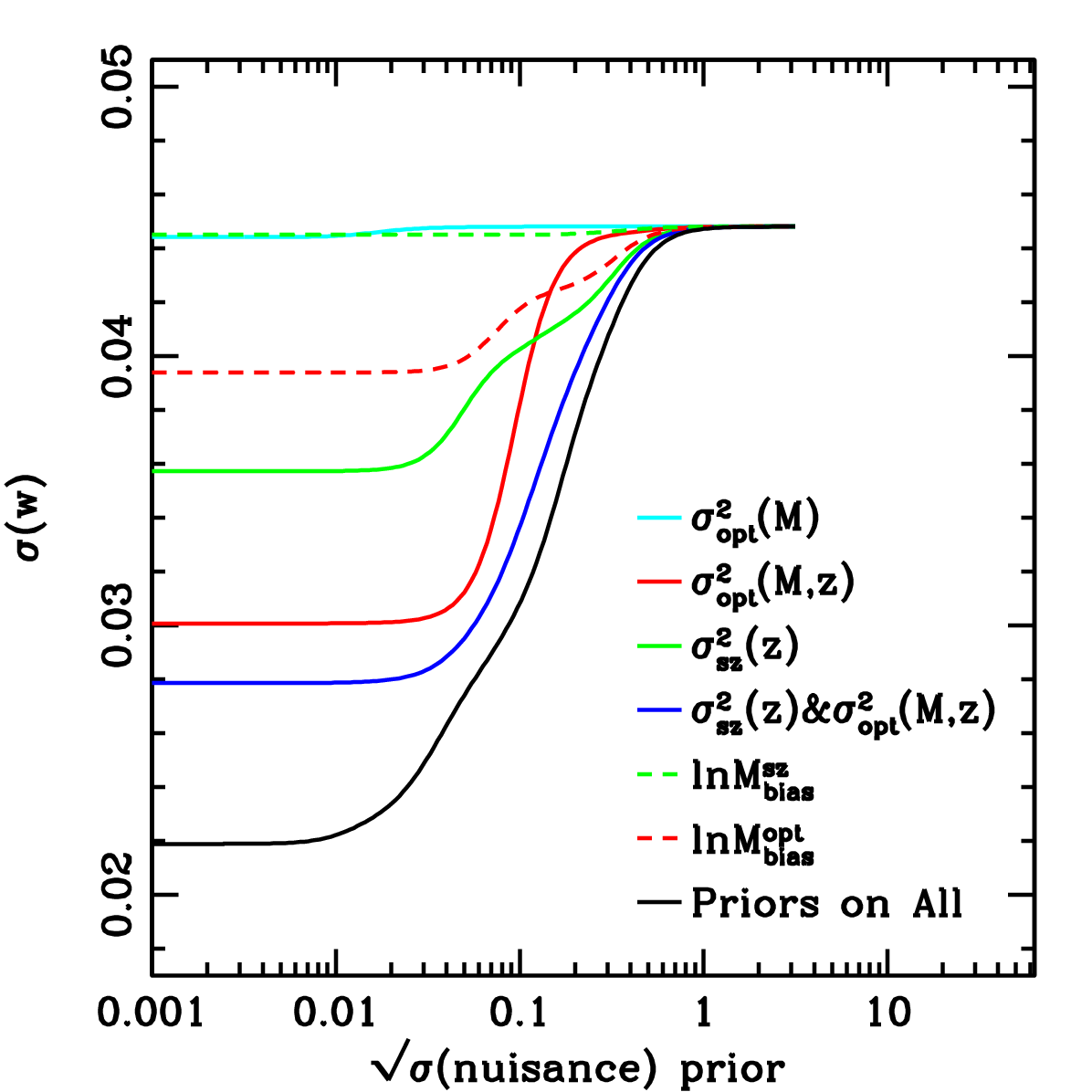}}
    \end{center}
  \end{minipage}
\end{center}
  \caption{$\sigma(\Omega_{DE})$({\it left}) and $\sigma(w)$ ({\it right}) {\it versus} 
the prior on the nuisance parameters for the full calibration analysis.
For the {\it cyan} lines, priors were applied on the mass dependent part of $\varopt$ only.
For the solid {\it red} lines priors were applied on all parameters of $\varopt$.
Applying priors to all terms of $\varsz$ yields the {\it solid  green} lines.
The {\it blue} lines were generated using priors on  $\varopt$ and $\varsz$.
The {\it dashed  green} lines have priors on ${\ln \Mbias_{\rm sz}}$ and the
 {\it dashed red} lines have priors on ${\ln \Mbias_{\rm opt}}$.
Applying priors to all nuisance parameters yields the {\it black} lines.
  }\label{fig:sig.pri}
  \end{figure*}

In Figure \ref{fig:sig.pri} we show $\sigma(\Omega_{DE})$ ({\it left}) and 
$\sigma(w)$  ({\it right}) as functions of the prior on the nuisance parameters 
for the full calibration analysis.
Throughout we assume that 
$\sigma_{\rm prior}=\sigma_{\rm prior}(\sigma^2_0)=\sigma_{\rm prior}(\ln \Mbias_0)=0.5\sigma_{\rm prior}(a_i)=0.5\sigma_{\rm prior}(b_i)=0.5\sigma_{\rm prior}(c_i)$.
We see from the {\it left} plot that constraints on $\Omega_{\rm DE}$ are most 
sensitive to priors on the mass bias, especially the optical mass bias. 
A prior of $(0.1)^2$ on ${\ln \Mbias_{\rm opt}}$ improves $\sigma(\Omega_{\rm DE})$ 
by a factor of $\sim3$.
With priors of $(0.1)^2$ on all parameters (multiplied by two where appropriate) 
$\sigma(\Omega_{\rm DE})$ improves by approximately an order of magnitude!

Constraints on $w$ are largely insensitive to priors on the mass-dependent part of the
optical scatter, $\sigma^2_{opt}(M)$, or on the SZ mass bias parameters.
Priors on the optical mass bias improve constraints by at most $12\%$.
The constraints are most sensitive to priors on the redshift dependent scatter
nuisance parameters, particularly the optical scatter.
A prior of $(0.1)^2$ on $\sigma^2_{opt}(M,z)$ and $\sigma^2_{SZ}(z)$ decreases
$\sigma(w)$ by a factor of $\sim 1.3$.
The full cross-calibration can constrain the constant parts of both the SZ and optical scatter 
so that priors on them do not improve $w$ constraints.    
The full improvement requires priors of  $(0.01)^2$ on all parameters 
and yields $\sigma(w)=0.022$.

\section{Conclusions and future work}\label{sec:conc}

We developed a formalism to derive joint cosmological and cluster mass-observable 
constraints from cluster number counts and clustering sample variance of multiple 
cluster finding techniques.
The improvement we find relative to previous works arises from our use of the 
interdependence of cluster measurements performed over the same patch of sky
to cross-calibrate the mass-observable relations of the different techniques.
When combining an SPT-like and DES-like survey, the full 
cross-calibration method yields $\sim 2$ times smaller constraints on
$\DE$ and $w$ compared to simply adding the Fisher matrices of the individual experiments.
Furthermore, constraints from the full cross-calibration are less sensitive 
to $\Mth$ and $z_{\rm max}$ than the single mass-observable constraints.

The cross-calibration places tight constraints on the correlation between the
observables without the need of additional priors.
Conversely, priors on the mass-variance and bias can significantly improve 
the dark energy constraints.
Constraints on $\DE$ are most sensitive to priors on the mass biases. 
On the other hand, constraints on $w$ are more sensitive to priors on the 
redshift-dependent part of the scatters.
Priors on the optical nuisance parameters are more relevant than priors on SZ
nuisance parameters for both $\DE$ and $w$ constraints.

Our technique can still be improved.
Combining more than two techniques at a time should further improve constraints.
But we can only combine multiple techniques if we use a more efficient 
binning strategy, to minimizes the number of mass bins needed to extract 
the useful information.
It is possible that a more efficient binning may improve even the two observable case,
particularly in cosmologies with low $\sigma_8$.

Work still needs to be done before the self-calibration or full cross-calibration 
can be applied to real data. 
The cross-calibration estimates presented here are sensitive to the parametrization of the 
mass errors.
Simulations are needed to determine what parametrizations are robust to theoretical
and experimental uncertainties.
Our results assumed a perfect selection, but selection effects may bias the cosmological 
constraints.
\cite{wu08} have shown that if the halo selection depends on halo concentration, and if the 
halo bias depends on the assembly history, the sample 
variance due to clustering will deviate from that of a random selection of halos with the 
same mass distribution.
If the clustering sample variance is modeled incorrectly, the self-calibration may bias
the recovered dark energy parameters. 
Since the different cluster surveys are expected to have selections with different 
dependence on the halo concentration, cross-calibration should mitigate
selection effects, though we are yet to test this hypothesis.
Finally, we must still account for the relation between photo-z and mass-observable errors.
Regardless of the simplifications adopted here, we conclude that having overlap between 
surveys is very important to maximize the effectiveness of cross-calibration techniques.
\acknowledgments
I would like to thank Marcos Lima for showing me how to self-calibrate and Dragan Huterer
for detailed comments on the text.
I would also like to thank Gus Evrard, Josh Frieman, Mike Gladders, Wayne Hu, Tim McKay, 
Stephan Meyer, Angela Olinto, Hiroaki Oyaizu, and Eduardo Rozo for useful discussions 
and helpful comments.
Some of the simulations used in this work have been performed on the Joint Fermilab - KICP 
Supercomputing Cluster, supported by grants from Fermilab, Kavli Institute for Cosmological 
Physics, and the University of Chicago.
This work was supported in part by the Kavli Institute for Cosmological Physics at the 
University of Chicago through grants NSF PHY-0114422 and NSF PHY-0551142 and an endowment 
from the Kavli Foundation and its founder Fred Kavli.

\appendix

\section{The probability distribution of multiple observables} \label{appendix}

Studies of the cluster mass-observable relation in the literature 
(e.g. \cite{ryk08a,joh07,rey08}), using either simulations or observations, 
typically estimate $p({\Mobs}| M)$ (by measuring the scatter of $\Mobs(M)$) for 
a single mass-observable or the relation between two observables, $p({\Mobs}_a|\Mobs_b)$, for a given $M$, or equivalently, assuming no evolution in $M$.
Thus, it is useful to express  $p({\bf \Mobs}| M)$ in terms of combinations of $p({\Mobs}| M)$ 
and $p({\Mobs}_a|\Mobs_b)$.
This can be done using the product rule of probability and Bayes' theorem. 
For example, for two observables,
\begin{eqnarray}
p({\bf \Mobs}| M)&=&p(\Mobs_a,\Mobs_b|M) \nonumber\\
&=&p(\Mobs_a|M)p(\Mobs_b|\Mobs_a,M) \nonumber\\
&=&p(\Mobs_a|M)p(\Mobs_b|M)\frac{p(\Mobs_a|\Mobs_b)}{p(\Mobs_a)}\nonumber \\
\end{eqnarray}

For  $n$ observables,
\begin{eqnarray}
p({\bf \Mobs}| M)=  \prod_{j=1}^{n-1} \left [\frac{\prod_{i=j+1}^{n-1} p(\Mobs_j|\Mobs_i)}{p(\Mobs_j)^{n-j}} \right] \nonumber \\
\times \prod_{i=1}^{n} p(\Mobs_j|M)
\end{eqnarray}

In this paper we focus on combining two observables at a time.
Given mass measurement techniques $a$ and $b$ we adopt the following parametrizations:

\begin{equation}
p(\Mobs_{o} | M) = \frac{1}{\sqrt{2\pi {\sigma}_{o}^{2}}} \exp\left[\frac{ -{x}_{o}^2(\Mobs_{o})}{2 \sigma_{o}^2} \right] ,
\end{equation}

\noindent where $o$ is either $a$ or $b$ and 
\begin{equation}
x_{o}(\Mobs_{o}) \equiv { \ln \Mobs_{o} - \ln M - \ln\Mbias_{o}} .
\end{equation}
\noindent The definition of $x_{o}(\Mobs_{o})$ here differs from the definition of $x(\Mobs)$ 
in Eq. (\ref{eqn:x1mobs}) by a factor of $\sqrt{2 \siglnM^2}$.

Similarly,

\begin{equation}
p(\Mobs_{a} |\Mobs_{b}) = {1 \over \sqrt{2\pi {\sigma}_{ab}^2} } \exp\left[ \frac{-x_{ab}^2(\Mobs_{ab})}{2{\sigma}_{ab}^2} \right] ,
\end{equation}
\noindent where 
\begin{eqnarray}
x_{ab}(\Mobs_{ab}) &\equiv& \ln \Mobs_{a} - \ln\Mbias_{a} - \ln \Mobs_{b} + \ln\Mbias_{b} \nonumber \\
&=& x_{a}-x_{b}
\end{eqnarray}
 
Combining all the probability distributions above, yields
\begin{equation}
p(\Mobs_a,\Mobs_b|M)=\frac{1}{\sqrt{8\pi^3 \sigma_{a}^2\sigma_{b}^2\sigma_{ab}^2}}\exp[A],
\end{equation}

\noindent where
\begin{equation}
A=\left[
-\frac{x_{a}^2}{2 \sigma_{a}^2}
-\frac{x_{b}^2}{2 \sigma_{b}^2}
-\frac{({x}_{a}-{x}_{b})^2}{2 \sigma_{ab}^2}
 \right] \label{eqn:A}
\end{equation}
\noindent and we have simplified the notation by writing $\sigma_{x}$ to represent ${\siglnM}_{x}$.
Rearranging the terms in \ref{eqn:A} we find
\begin{eqnarray}
A&=&\frac{-1}{ 2}\left[ {x}_{a}^2\left( \frac{1}{\sigma_{a}^2} +  \frac{1}{\sigma_{ab}^2} \right) + {x}_{b}^2\left( \frac{1}{\sigma_{b}^2} +  \frac{1}{\sigma_{ab}^2} \right) \right.  \nonumber \\
&& \left. - 2{x}_{a}{x}_{b}\left(\frac{1}{{\sigma_{ab}^2}} \right)\right]
\end{eqnarray}

\noindent If we define the vector ${\bf x}=(x_{a},x_{b})$ and the matrix 
\begin{equation}
{\bf B}=\begin{pmatrix}
\frac{1}{{{\sigma}_{a}^2}} + \frac{1}{{{\sigma}_{ab}^2}} & - {1\over{\sigma}_{ab}^2} \\
- {1\over{\sigma}_{ab}^2} & \frac{1}{{{\sigma}_{a}^2}} +  \frac{1}{{{\sigma}_{ab}^2}}
\end{pmatrix}
\end{equation}
\noindent we obtain
\begin{equation}
A=\frac{-1}{2} \left[
{\bf x}^{\rm T} {\bf B} {\bf x}
\right]
\end{equation}
With the above form for $A$, it is clear that we can represent $p(\Mobs_1,\Mobs_2|M)$
by a bivariate Gaussian distribution,
\begin{equation}
p(\Mobs_1,\Mobs_2|M)={1 \over (2\pi)\det({\bf C})^{1/2}} \exp\left[ -{\bf x}^{\rm T} {\bf C}^{-1} {\bf x}\right]
\end{equation}
\noindent where $\bf C$ is the covariance matrix defined as
\begin{equation}
{\bf C}=\begin{pmatrix}
{\sigma}_{a}^2 & \rho {\sigma}_{a}{\sigma}_{b}\\
\rho {\sigma}_{a}{\sigma}_{b} & {\sigma}_{b}^2
\end{pmatrix}
\end{equation}
\noindent and $\rho$ is the correlation coefficient defined in terms of ${\sigma}_{a}$, ${\sigma}_{b}$, and ${\sigma}_{ab}$ as 
\begin{equation}
\rho = \frac{{\pm {\sigma}_{a}{\sigma}_{b}}}{{ \left[ ({\sigma}_{a}^2+{\sigma}_{ab}^2 )({\sigma}_{b}^2+{\sigma}_{ab}^2)   \right]^{1/2} }} \label{eqn:rhosig}
\end{equation}

\bibliography{thesis.prd}
\end{document}